\documentclass[12pt]{article}

\usepackage{amsthm,amsmath,amsfonts}
\usepackage{latexsym}
\usepackage[T1]{fontenc}
\usepackage{a4wide}
\usepackage{times}
\usepackage{array}
\usepackage{hhline}
\usepackage{graphicx}

\numberwithin{equation}{section}

\newtheorem{tw}{Theorem}
\newtheorem{lem}{Lemma}

\title{On an inhomogeneous slip-inflow boundary value problem for a steady flow of a viscous compressible fluid
	in a cylindrical domain}
\author{Tomasz Piasecki}

\date{\today}

\begin{document}

\maketitle

\centerline{\bf Abstract}

We investigate a steady flow of a viscous compressible fluid with inflow boundary condition on the density
and inhomogeneous slip boundary conditions on the velocity in a cylindrical domain $\Omega = \Omega_0 \times (0,L) \in \mathbb{R}^3$.
We show existence of a solution $(v,\rho) \in W^2_p(\Omega) \times W^1_p(\Omega)$,
where $v$ is the velocity of the fluid and $\rho$ is the density, that is a small perturbation
of a constant flow $(\bar v \equiv [1,0,0], \bar \rho \equiv 1)$.
We also show that this solution is unique in a class of small perturbations of $(\bar v,\bar \rho)$.
The term $u \cdot \nabla w$ in the continuity equation makes it impossible to show the existence
applying directly a fixed point method.
Thus in order to show existence of the solution we construct a sequence $(v^n,\rho^n)$
that is bounded in $W^2_p(\Omega) \times W^1_p(\Omega)$ and satisfies the Cauchy condition in
a larger space $L_{\infty}(0,L;L_2(\Omega_0))$ what enables us to deduce that the weak limit of a subsequence of
$(v^n,\rho^n)$ is in fact a strong solution to our problem.

\section{Introduction}
The mathematical description of a flow of a viscous, compressible fluid usually lead to problems
of mixed character as the momentum equation is elliptic (in stationary case) or parabolic
(in case of time-dependent flow) in the velocity, while the continuity equation is hyperbolic in the density.
Therefore, the application of standard methods usually applied to elliptic or hyperbolic problems fails
in the mathematical analysis of the compressible flows and a combination of such techniques, as well
as development of new mathematical tools is required. As a result a consistent theory of weak solutions
to the Navier - Stokes equations for compressible fluids has been developed quite recently
in the 90's, mainly due to the work of Lions \cite{PL} and Feireisl \cite{Fe}.
An overview of these results is given in
the monograph \cite{NoS}. A modification of this approach in case of steady flows with slip boundary
conditions has been developed
by Mucha and Pokorny in a dwo dimensional case in \cite{PMMP1} and in 3D in \cite{PMMP2}.

The issue of regular solutions is less investigated and the problems are considered mainly with Dirichlet
boundary conditions. If we assume that the velocity does not vanish on the boundary, the hyperbolicity
of the continuity equation makes it necessary to prescribe the density on the part of the boundary
where the flow enters the domain.
In \cite{VZ} Valli and Zajaczkowski investigate a time-dependent system with inflow boundary condition,
obtaining also a result on existence of a solution to stationary problem.
The existence of regular solutions to stationary problems with an inflow conditon on the density
has been investigated by Kellogg and Kweon \cite{Kw1} and Kweon and Song \cite{Kw3}.
Their results require some smallness assumptions on the data, and the regularity of solutions is a subject
to some constraints on the geometry of the boundary near the points where the inflow and outlow parts of the boundary
meet. In \cite{Kw2} Kellogg and Kweon consider a domain where the inflow and outflow parts of the
boundary are separated, obtaining regular solutions.

The lack of general existence results inhibits the development of qualitative analysis of compressible flows.
Therefore it is worth to mention here the papers by Plotnikov and Sokolowski who has investigated shape optimization
problems with inflow boundary condition in 2D \cite{PS2} and 3D \cite{PS3} dealing with weak solutions.
More recently Plotnikov, Ruban and Sokolowski have investigated shape optimization problems working with
strong solutions in \cite{PRS1} and \cite{PRS2}.

It seems interesting both from the mathematical point of view and in the eye of applications
to investigate problems with inflow boundary condition on the density combined with slip boundary
conditions on the velocity, that enables to describe precisely the action between the fluid and
the boundary. Such problem is investigated in this paper. The domain is a three dimensional cylinder
and we assume that the fluid slips along the boundary with a given friction coefficient and
there is no flow across the wall of the cylinder. We show existence of a regular solution
that can be considered a small perturbation of a constant solution. The method of the proof
is outlined in the next part of the introduction and now we are in a position to formulate
our problem more precisely.

\begin{figure}[htb]
\begin{center}
\includegraphics[width= 0.7 \textwidth]{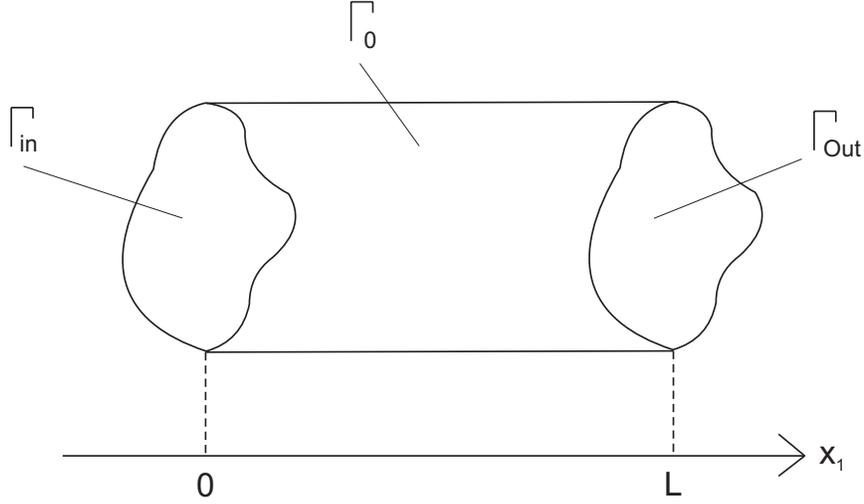}
\caption{The domain}      \label{rys1}
\end{center}
\end{figure}

The flow is described by the
Navier-Stokes system supplied with the slip boundary conditions on the velocity.
The complete system reads
\begin{eqnarray}  \label{main_system}
\begin{array}{lcr}
\rho v \cdot \nabla v -\mu \Delta v - (\mu+\nu) \nabla {\rm div}\, v
+\nabla \pi(\rho) =0 & \mbox{in} & \Omega,\\
{\rm div}\,(\rho v)=0 & \mbox{in} & \Omega,\\
%\end{array}
%\end{eqnarray}
%$$
%
%$$
%\begin{array}{lcr}
n\cdot {\bf T}(v,\pi(\rho))\cdot \tau_k +f v\cdot \tau_k=b_k, \quad k=1,2 &
\mbox{on} &\Gamma,\\
n \cdot v=d & \mbox{on} & \Gamma,\\
\rho=\rho_{in} & \mbox{on} & \Gamma_{in},
\end{array}
\end{eqnarray}
where $v: \mathbb{R}^3 \to \mathbb{R}^3$ is the unknown velocity field of the fluid and $\rho:\mathbb{R}^3 \to \mathbb{R}$
is the unknown density. We assume that the pressure is a function of the density of a class $C^3$.
Further, $\mu$ and $\nu$ are viscosity coefficients satisfying $(\mu + 2\nu)>0$ and $f>0$ is a friction coefficient.
The domain $\Omega$ is a cylinder in $\mathbb{R}^3$ of a form
$\Omega = \Omega_0 \times (0,L)$ where $\Omega_0 \in \mathbb{R}^2$
is a set with a boundary regular enough and $L$ is a positive constant (see fig. \ref{rys1}).
%
%\begin{figure}[htb]
%\begin{center}
%\includegraphics[width= 0.7 \textwidth]{cylinder_rys1.jpg}
%\caption{The domain}      \label{rys1}
%\end{center}
%\end{figure}
%
We want to show existence of a solution that can be considered a small perturbation of a constant
flow $(\bar v, \bar \rho) \equiv ([1,0,0],1)$. Thus we denote the subsets of the boundary
$\Gamma = \partial \Omega$ as $\Gamma = \Gamma_{in} \cup \Gamma_{out} \cup \Gamma_0$,
where $\Gamma_{in} = \{ x \in \Gamma: \bar v \cdot n < 0 \} $,
$\Gamma_{out} = \{ x \in \Gamma: \bar v \cdot n > 0 \}$ and $\Gamma_{0} = \{ x \in \Gamma: \bar v \cdot n = 0 \}$.

By $n$ we denote the outward unit normal to $\Gamma$ and $\tau_1, \tau_2$ are the unit tangent vectors
to $\Gamma$.
Since the boundary has singularities at the junctions of $\Gamma_{in}$ and $\Gamma_{out}$
with $\Gamma_0$, for the boundary traces we will consider functional spaces that are
algebraic sums of spaces defined on the boundary. More precisely for $s,q \in \mathbb{R}$ we shall denote
$
W^q_s(\Gamma) := W^q_s(\Gamma_{in}) + W^q_s(\Gamma_{out}) + W^q_s(\Gamma_{0}).
$
We assume that $b \in W^{1-1/p}_p(\Gamma)$, $\rho_{in} \in W^{1}_p(\Gamma_{in})$
and $d \in W^{2-1/p}_p(\Gamma)$ are given functions and
$d=0$ on $\Gamma_0$ what means that $\Gamma_0$ is an impermeable wall.

For simplicity we consider the momentum equation with zero r.h.s., but our proofs work without
any modification for the r.h.s. $\rho \, F$ where $F$ is small enough in $L_p$.

We shall make here some remarks concerning notation. Since we will usually use the spaces of functions
defined on $\Omega$, we will skip $\Omega$ in notation of the spaces, for example we will write
$L_2$ instead of $L_2(\Omega)$. For the density we will use estimates in the space
$L_{\infty}(0,L;L_2(\Omega_0))$. For simplicity we will denote this space by $L_{\infty}(L_2)$.
A constant dependent on the data that can be controlled, but not necessarily
small, will be denoted by $C$, and $E$ shall denote a constant that can be arbitrarily small provided
that the data is small enough.

In order to formulate our main result let us define a quantity $D_0$ that measures how the boundary data
$b$,$d$ and $\rho_{in}$ differ from the values of, respectively, $f \bar v \cdot \tau_i$, $n \cdot \bar v$
and $\bar \rho$ in appropriate norms.
We have $\bar v \cdot \tau_i = \tau_i^{(1)}$ and $\bar v \cdot n = n^{(1)}$,
thus we define
\begin{equation} \label{D0}
D_0 = ||b_i - f \tau_i^{(1)}||_{W^{1-1/p}_p(\Gamma)} + ||d - n^{(1)}||_{W^{2-1/p}_p(\Gamma)}
+ ||\rho_{in}-1||_{W^{1}_p(\Gamma_{in})}.
\end{equation}
Our main result is
\begin{tw}  \label{main}
Assume that $D_0$ defined in (\ref{D0}) is small enough, $f$ is large enough
and $p>3$.
Then there exists a solution $(v,\rho) \in W^2_p(\Omega) \times W^1_p(\Omega)$
to the system (\ref{main_system}) and
\begin{equation} \label{est_main}
||v - \bar v||_{W^2_p} + ||\rho - \bar \rho||_{W^1_p} \leq E(D_0),
\end{equation}
where $E(D_0)$ can be arbitrarily small provided that $D_0$ is small enough.
This solution in unique in the class of solutions satisfying the estimate (\ref{est_main}).
\end{tw}
The major difficulty in the proof of Theorem \ref{main} is in the term $u \cdot \nabla w$
in the continuity equation, that yields impossible a direct application of a fixed point argument.
To overcome this problem one can apply the method of elliptic regularization, known rather from the theory
of weak solutions (see \cite{NoS}). This method has been applied to a similar problem in a two dimensional case
in \cite{TP2}. However, it complicates considerably the computations since we have to find the bound on the artificial
diffusive term. 
Here we apply a method of successive approximations, that leads to a more direct proof.  
In order to prove Theorem \ref{main} we will construct a sequence $(u^n,w^n) \in W^2_p \times W^1_p$
that converges to the solution of (\ref{main_system}). Due to the presence of the term
$u \cdot \nabla w$ we can not show directly the convergence in $W^2_p \times W^1_p$,
but we can show that $(u^n,w^n)$ is a Cauchy sequence in a larger space $H^1 \times L_{\infty}(L_2)$
and thus converges in this space to the weak solution of (\ref{main_system}).
On the other hand, the sequence will converge on a subsequence weakly in $W^2_p \times W^1_p$,
what will enable us to show that
the weak solution is in fact strong. A similar approach has been applied
in \cite{MD} to an evolutionary Navier-Stokes system in a framework of Besov spaces.

We start with removing the inhomogeneity from the boundary condition (\ref{main_system})$_{4}$.
To this end let us construct
$u_0 \in W^2_p(\Omega)$ such that
\begin{equation}  \label{extension}
n\cdot u_0|_{\Gamma}= d - n^{(1)}.
\end{equation}
Due to the assumption of smallness of $d - n^{(1)}$ in $W^{2-1/p}_p(\Gamma)$ we can assume
that
\begin{equation} \label{small}
||u_0||_{W^2_p} << 1.
\end{equation}
From now on we assume (\ref{small}) in all our results. Now we consider
$$
u = v - \bar v - u_0\; \mbox{ \ \ \
and \ \ \ } w = \rho - \bar \rho.
$$
One can easily verify that $(u,w)$ satisfies the following system:
\begin{eqnarray} \label{system}
\begin{array}{lcr}
\partial_{x_1} u -\mu \Delta  u - (\nu + \mu) \nabla {\rm div}\, u +
\pi'(1) \, \nabla  w =  F(u,w) & \mbox{in} & \Omega,\\
{\rm div}\, u + \partial_{x_1}w + (u+u_0) \cdot \nabla  w = G(u,w)
& \mbox{in}& \Omega,\\
n\cdot 2\mu {\bf D}( u)\cdot \tau_i +f \ u \cdot \tau_i = B_i, \quad i=1,2
&\mbox{on} & \Gamma, \\
n\cdot  u = 0 & \mbox{on} & \Gamma,\\
w=w_{in} & \mbox{on} & \Gamma_{in},\\
\end{array}
\end{eqnarray}
where
\begin{equation}  \label{FG}
\begin{array}{c}
F(u,w) = - w \, (u + \bar v + u_0) \cdot \nabla (u+u_0) - (u_0 \cdot \nabla u) - u \cdot \nabla u_0
\\
+ \mu \Delta u_0 + (\nu+\mu) \nabla {\rm div}\, u_0 - u_0 \cdot \nabla u_0
- [\pi'(w+1) - \pi'(1)] \nabla w,
\\
G(u,w) = -(w + 1) \, {\rm div}\, u_0 - w \, {\rm div}\,u
\end{array}
\end{equation}
and
$$ 
B_i = b_i - 2 \mu \, n \cdot \mathbf{D}(u_0) \cdot \tau_i - f \tau_i^{(1)}.
$$
From now on we will denote $\pi'(1)=:\gamma$.
We see that $F$ and $G$ also depend on $\nabla u,u_0,\nabla u_0$, but for simplicity
we will write $F(u,w)$ and $G(u,w)$.
In order to prove Theorem \ref{main} it is enough to show the existence of a solution $(u,w)$
to the system (\ref{system}) provided that $||B||_{W^{1-1/p}_p(\Gamma)}$ and $||u_0||_{W^2_p(\Omega)}$ are small enough.
As we already mentioned, we will construct a sequence that converges to the solution.
The sequence will be defined as
\begin{eqnarray} \label{system_seq}
\begin{array}{lcr}
\partial_{x_1} u^{n+1} -\mu \Delta u^{n+1} - (\nu + \mu) \nabla {\rm div}\, u^{n+1} +
\gamma \nabla  w^{n+1} =  F(u^n,w^n) & \mbox{in} & \Omega,\\
{\rm div}\, u^{n+1} + \partial_{x_1}w^{n+1} + (u^n+u_0) \cdot \nabla  w^{n+1} = G(u^n,w^n)
& \mbox{in}& \Omega,\\
n\cdot 2\mu {\bf D}(u^{n+1})\cdot \tau_i +f \ u^{n+1} \cdot \tau_i = B_i, \quad i=1,2
&\mbox{on} & \Gamma, \\
n\cdot  u^{n+1} = 0 & \mbox{on} & \Gamma,\\
w^{n+1}=w_{in} & \mbox{on} & \Gamma_{in}.\\
\end{array}
\end{eqnarray}
As we will see in the sequel, our method does not require any particular starting point for the sequence $(u^n,w^n)$,
but only some smallness assumptions on the starting point $(u^0,w^0)$,
hence without loss of generality we can set
$(u^0,w^0) = (0,0)$.
In order to show the existence of the sequence defined in (\ref{system_seq}) we have to solve
a linear system:
\begin{eqnarray} \label{system_lin}
\begin{array}{lcr}
\partial_{x_1} u -\mu \Delta  u - (\nu + \mu) \nabla {\rm div} \, u +
\gamma \nabla  w =  F & \mbox{in} & \Omega,\\
{\rm div}\, u + \partial_{x_1}w + (\bar u + u_0) \cdot \nabla  w = G
& \mbox{in}& \Omega,\\
n\cdot 2\mu {\bf D}( u)\cdot \tau_i +f \ u \cdot \tau_i = B_i, \quad i=1,2
&\mbox{on} & \Gamma, \\
n\cdot  u = 0 & \mbox{on} & \Gamma,\\
w=w_{in} & \mbox{on} & \Gamma_{in},\\
\end{array}
\end{eqnarray}
where $(F,G,\bar u,u_0) \in L_p \times W^1_p \times W^2_p \times W^2_p$
are given functions and $\bar u \cdot n = 0$ on $\Gamma$.

Let us now outline the strategy of the proof, and thus the structure of the paper.
In section \ref{sec_apriori} we show the \emph{a priori} estimate (\ref{est_lin_w2p})
on a solution to the linear system (\ref{system_lin}).
We start with an energy estimate in $H^1 \times L_{\infty}(L_2)$. Next the properties of the slip boundary conditions
enables us to show that the vorticity of the velocity on the boundary has the same regularity as the velocity,
and this fact makes it possible to find a bound on $||w||_{W^1_p}$. Then the estimate (\ref{est_lin_w2p}) results
directly from the elliptic regularity of the Lame system.

The linear system (\ref{system_lin}) is solved in section \ref{sec_sol_lin}.
First we show the existence of a weak solution
using the Galerkin method modified to deal with the continuity equation.  
Next we can show
that this solution is in fact strong using \emph{a priori} estimate and symmetry of the slip boundary conditions.

In section \ref{sec_conv} we show the estimate in $W^2_p \times W^1_p$ on the sequence $(u^n,w^n)$
and, as a result, the Cauchy condition satisfied by this sequence in the space $H^1 \times L_{\infty}(L_2)$.
These results are derived by application of the estimates for the linear system.

In section \ref{sec_proof} we apply the results of section \ref{sec_conv} passing to the limit with  
$(u^n,w^n)$ and then showing that the limit
is a solution to (\ref{system}). Finally we show that this solution is unique in a class of solutions
satisfying the estimate (\ref{est_main}).

\section{A priori bounds}  \label{sec_apriori}
The main result of this section is the estimate (\ref{est_lin_w2p}) in $W^2_p \times W^1_p$.
In order to show it we start with an energy estimate in $H^1 \times L_\infty(L_2)$. Next we consider the
equation on the vorticity of the velocity and apply the Helmholtz decomposition to derive the bound
on $||w||_{W^1_p}$ and finally using the classical elliptic theory we conclude (\ref{est_lin_w2p}).

In our proofs we shall not need explicit formulas on the functions $F(u,w)$ and $G(u,w)$, what will
be important is that they depend quadratically on $u$ and $w$.
More precisely, we will show a following estimate
\begin{lem}
Let $(u,w) \in W^2_p \times W^1_p$ and let $F(u,w)$ and $G(u,w)$ be defined in (\ref{FG}). Then
\begin{equation}  \label{FGlp}
\begin{array}{c}
||F(u,w)||_{L_p} + ||G(u,w)||_{W^1_p} \leq \\
\leq C \, [ (||u||_{W^2_p} + ||w||_{W^1_p})^2 + ||u_0||_{W^2_p} ].
\end{array}
\end{equation}
\end{lem}
\textbf{Proof} Since by the imbedding theorem $W^1_p(\Omega) \subset L_{\infty}(\Omega)$,
the estimate on $||G||_{W^1_p}$ is straightforward, and the only part of $F$ that deserves attention is
$\delta \pi'(w) \nabla w$, where 
\begin{equation} \label{delta_pi}
\delta \pi'(w) := \pi'(w+1) - \pi'(1).
\end{equation}
We will apply a fact that for a $C^1$ - function $f$ we have
\begin{equation}  \label{mean}
f(x) - f(y) = (x-y) \, \int_{0}^1 f'[tx + (1-t)y] \,dt,
\end{equation}
Thus we have
$$
\delta \pi'(w) = w \, \int_0^1 \pi''(tw + 1) \,dt.
$$
Since $\pi$ is a $C^3$ - function, the above implies
$$
||\delta \pi'(w) \nabla w||_{L_p} \leq C(\pi) ||w||_{\infty} ||\nabla w||_{L_p} \leq C \, ||w||_{W^1_p}^2.
$$
The other parts of $F$ can be estimated direcly giving (\ref{FGlp}). $\square$

Next, we derive the 'energy' estimate in $H^1 \times L_{\infty}(L_2)$. It is stated in the following lemma
\begin{lem}
Let $(u,w)$ be a solution to the system (\ref{system_lin}) 
with $(F,G,B,w_{in}, \bar u) \in V^* \times L_2 \times L_2(\Gamma) \times L_2(\Gamma_{in}) \times W^2_p$, with
$||\bar u||_{W^2_p}$ small enough and $f$ large enough. Then
\begin{equation} \label{ene1}
||u||_{H^1} + ||w||_{L_{\infty}(L_2)} \leq C \, [||F||_{V^*} + ||G||_{L_2} + ||B||_{L_2(\Gamma)} + ||w_{in}||_{L_2(\Gamma_{in})}],
\end{equation}
where
\begin{equation} \label{def_V}
V = \{v \in H^1(\Omega): v \cdot n|_{\Gamma} = 0 \}
\end{equation}
and $V^*$ is the dual space of $V$.
\end{lem}
\textbf{Proof.}
We apply a general identity
\begin{equation} \label{basic_id}
\begin{array}{c}
\int_{\Omega} (-\mu \Delta u - (\nu + \mu) \nabla {\rm div}\,u) \, v \,dx =
\\
=\int_{\Omega} \{2 \mu \mathbf{D}(u) : \nabla \,v + \nu {\rm div} \, u \, {\rm div} \, v  \} \,dx
- \int_{\Gamma} n \cdot [ 2\mu \mathbf{D}(u) + \nu \, {\rm div}\,u \, \mathbf{Id} ] \cdot v \,d\sigma.
\end{array}
\end{equation}
For $u,v$ satisfying the boundary conditions (\ref{system_lin})$_{3,4}$ the boundary term in (\ref{basic_id}) equals
$$
\int_{\Gamma} \{ \sum_{i=1}^2 [B_i - f(u \cdot \tau_i)] (v \cdot \tau_i) \} \,d\sigma.
$$
Thus multiplying (\ref{system_lin})$_1$ by $u$ and integrating over $\Omega$ we get
\begin{equation}  \label{lem1_1}
\begin{array}{c}
\int_{\Omega} \{ 2\mu \mathbf{D}^2 (u) + \nu {\rm div}^2 \,u \} \,dx
+ \int_{\Gamma} (f + \frac{n^{(1)}}{2} ) |u|^2 \,d\sigma
- \gamma \int_{\Omega} w \, {\rm div}\,u \,dx = \\
= \int_{\Omega} F \cdot u \,dx + \int_{\Gamma} \{ B_1 (u \cdot \tau_1) + B_2 (u \cdot \tau_2) \} \,d\sigma.
\end{array}
\end{equation}
From now on (not only in this proof but also later)
we will use the summation convention when taking the sum over the tangential components.
Applying (\ref{system_lin})$_2$ and the boundary conditions we get
$$
\int_{\Omega} w {\rm div}\,u \,dx =
\int_{\Omega} Gw \,dx + \frac{1}{2} \int_{\Omega} w^2 \, {\rm div}\,(\bar u+u_0) \,dx
$$$$
- \frac{1}{2} \int_{\Gamma_{out}} w^2 (1 + u_0^{(1)}) \,d\sigma
+ \frac{1}{2} \int_{\Gamma_{in}} w_{in}^2 (1 + u_0^{(1)}) \,d\sigma.
$$
For $||u_0||_{W^2_p}$ small enough we have by the imbedding theorem
$1 + u_0^{(1)} > 0$ a.e. on $\Gamma_{out}$ what yields
$\int_{\Gamma_{out}} w^2 (1 + u_0^{(1)}) \, d\sigma > 0$. Moreover,
for the friction $f$ large enough on $\Gamma_{in}$ the boundary term
in (\ref{lem1_1}) will be positive. Combining these facts with the Korn inequality
(that can be proved in a simple way with the friction large enough - see Lemma 2.4 in \cite{TP1}): 
\begin{equation} \label{Korn}
\int_{\Omega} 2 \mu {\bf D}^2(u) + \int_{\Gamma} f (u \cdot \tau)^2 \,d\sigma \geq C \, ||u||_{H^1}^2
\end{equation}
we derive from (\ref{lem1_1}) the following inequality
\begin{equation} \label{lem1_2}
C ||u||_{H^1}^2 \leq \int_{\Omega} F \cdot u \,dx + \int_{\Gamma} B_i (u \cdot \tau_i) \,d\sigma
+ \frac{1}{2} \int_{\Omega} w^2 \, {\rm div}\, (\bar u+u_0) \,dx
- \frac{1}{2} \int_{\Gamma_{in}} w^2_{in} (1+u_0^{(1)}) \,d\sigma.
\end{equation}
In order to derive (\ref{ene1}) from (\ref{lem1_2}) we have to estimate $||w||_{L_{\infty}(L_2)}$
in terms of $||u||_{H^1}$ and the data. To show this estimate we refer to section \ref{sec_sol_lin}
where the linear system (\ref{system_lin}) is solved. Namely, we have $w = S(G - {\rm div}\,u)$
where the operator $S$ is defined in (\ref{S}) and thus the estimate (\ref{est_S}) implies
\begin{equation}
||w||_{L_{\infty}(L_2)} \leq C \, (||G||_{L_2} + ||u||_{H_1} + ||w_{in}||_{L_2(\Gamma_{in})}).
\end{equation}
The above inequality combined with (\ref{lem1_2}) yields (\ref{ene1}). $\square$

Now we consider the vorticity of the velocity $\alpha = {\rm rot}\,u$.
The properties of the slip boundary conditions enables us to express the tangential
components of $\alpha$ on the boundary in terms of the velocity. We arrive at the following
system
\begin{equation}   \label{system_rot}
\begin{array}{lcr}
\partial_{x_1} \alpha - \mu \Delta \alpha = {\rm rot}\,F & \mbox{in} & \Omega, \\
\alpha \cdot \tau_2 = (2 \chi_1 - \frac{f}{\nu}) u \cdot \tau_1 + \frac{B_1}{\nu} & \mbox{on} & \Gamma, \\
\alpha \cdot \tau_1 = (\frac{f}{\nu} - 2 \chi_2) u \cdot \tau_2 - \frac{B_2}{\nu} & \mbox{on} & \Gamma, \\
{\rm div}\,\alpha = 0 & \mbox{on} & \Gamma,
\end{array}
\end{equation}
where $\chi_i$ denote the curvatures of the curves generated by tangent vectors $\tau_i$.
In order to show the boundary relations (\ref{system_rot})$_{2,3}$ it is enough to differentiate
(\ref{system_lin})$_4$ with respect to the tangential directions and apply (\ref{system_lin})$_3$.
A rigorous proof, modifying the proof in the two-dimentional case from \cite{MR}, is given in the Appendix.
The condition ${\rm div}\, \alpha = 0$ in $\Omega$ results simply
from the fact that $\alpha = {\rm rot}\,u$. We introduce this relation as a boundary condition
(\ref{system_rot})$_4$, that completes
the conditions on the tangential parts of the vorticity. What is remarkable in
the boundary conditions (\ref{system_rot})$_{2,3}$ is that the tangential parts of
the vorticity on the boundary has the same regularity as the velocity itself and the data.
This feature of slip boundary conditions makes it possible to show the higher estimate on the vorticity
(see \cite{PM1},\cite{PM2}, \cite{PMMP2}).

In order to derive the bound on the vorticity we can follow \cite{PMMP2}, Lemma 4,
and construct $\alpha_0$, a divergence-free extension of the boundary data
$(\ref{system_rot})_{2,3}$, for example as a solution to the Stokes problem with zero r.h.s and
the boundary conditions $(\ref{system_rot})_{2,3}$ supplied with $\alpha_0 \cdot n = 0$.
The theory of the Stokes system then yields
\begin{equation} \label{alpha0_w1p}
||\alpha_0||_{W^1_p} \leq C \, \big[ ||u||_{W^{1-1/p}_p(\Gamma)} + ||B||_{W^{1-1/p}_p(\Gamma)} \big].
\end{equation}
Then the function $\alpha-\alpha_0$ satisfies the system
\begin{equation}
\begin{array}{lcr}
- \mu \Delta (\alpha-\alpha_0) = {\rm rot}\, [F - \partial_{x_1}u] + \mu \Delta \alpha_0 & \mbox{in} & \Omega, \\
(\alpha-\alpha_0) \cdot \tau_1 = 0 & \mbox{on} & \Gamma, \\
(\alpha-\alpha_0) \cdot \tau_2 = 0 & \mbox{on} & \Gamma, \\
{\rm div} \, (\alpha - \alpha_0) = 0 & \mbox{on} & \Gamma.
\end{array}
\end{equation}
Here we have used the fact that $\partial_{x_1} \alpha = {\rm rot} \partial_{x_1}u$ to preserve
the rotational structure of the r.h.s.
For the above system we have the following estimate (see \cite{Z})
\begin{equation} \label{alpha_w1p_1}
||\alpha||_{W^1_p} \leq C \, \big[||F||_{L_p} + ||\partial_{x_1}u||_{L_p} + ||\alpha_0||_{W^1_p} \big].
\end{equation}
The term with $\alpha_0$ can be bounded by (\ref{alpha0_w1p}) and to deal with $\partial_{x_1}u$
we apply the interpolation inequality (\ref{int1}). We obtain the term $||u||_{H^1}$ that we bound using
(\ref{ene1}) and finally arrive at
\begin{equation}  \label{rotuw1p}
||\alpha||_{W^1_p} \leq C(\epsilon) \, [ ||F||_{L_p} + ||G||_{W^1_p} +||w_{in}||_{L_2(\Gamma_{in})} + ||u||_{W^{1-1/p}_p(\Gamma)} + ||B||_{W^{1-1/p}_p(\Gamma)} ]
			+ \epsilon ||u||_{W^2_p}.
\end{equation}
%
%The system for $\alpha_2$ is standard once we deal with the edge singularity of the boundary,
%but this singularity can be easily removed using the symmetry of the domain. Namely, we can
%reflect $\Omega$ symmetricaly w.r.t. $\Gamma_0$ and on the reflected part define
%a vector field $\tilde \alpha_2$:
%
%\begin{displaymath}
%\begin{array}{c}
%\tilde \alpha_2^{(1)}(-x_1,x_2,x_3) = \alpha_2^{(1)}(x_1,x_2,x_3), \\
%\tilde \alpha_2^{(2)}(-x_1,x_2,x_3) = -\alpha_2^{(2)}(x_1,x_2,x_3), \\
%\tilde \alpha_2^{(3)}(-x_1,x_2,x_3) = -\alpha_2^{(3)}(x_1,x_2,x_3).
%\end{array}
%\end{displaymath}
%
%The transformation preserves the boundary conditions (\ref{system_alpha2})$_{2,3,4}$
%and we can proceed solving the equation in the extended domain to obtain $\alpha_2$
%a solution to (\ref{system_alpha2}) satisfying
%
With the bound on the vorticity at hand the next step is to consider
the Helmholtz decomposition of the velocity (the proof can be found in \cite{Ga}):
\begin{equation} \label{Helm}
u = \nabla \phi + A,
\end{equation}
where $\phi|_{\Gamma}=0$ and ${\rm div}\, A =0$.
We see that the field $A$ satisfies the following system
\begin{equation}
\begin{array}{lcr}
{\rm rot}\, A = \alpha & \mbox{in} & \Omega, \\
{\rm div}\, A = 0 & \mbox{in} & \Omega, \\
A \cdot n = 0 & \mbox{on} & \Gamma.
\end{array}
\end{equation}
This is the standard rot-div system and we have
$
||A||_{W^2_p} \leq C \, ||\alpha||_{W^1_p} ,
$
what by (\ref{rotuw1p}) can be rewritten as
\begin{equation}  \label{Aw2p}
||A||_{W^2_p} \leq C(\epsilon) \, [ ||F||_{L_p} + ||G||_{W^1_p} + ||u||_{W^{1-1/p}_p(\Gamma)} + ||B||_{W^{1-1/p}_p(\Gamma)}
		+ ||w_{in}||_{W^1_p(\Gamma_{in})} ] + \epsilon ||u||_{W^2_p}
\end{equation}
for any $\epsilon>0$.
Now we substitute the Helmholtz decomposition to (\ref{system_lin})$_1$. We get
\begin{equation}  \label{nablaH}
\nabla [ -(\nu + 2\mu) \Delta \phi + \gamma \, w ] =
F - \partial_{x_1} A + \mu \Delta A + (\nu+\mu) \nabla \,{\rm div}\,A - \partial_{x_1} \phi,
\end{equation}
but $\Delta \phi = {\rm div}\, u$ and denoting the l.h.s. of the above equation by $\bar F$ we obtain
\begin{equation}
-(\nu + 2\mu) {\rm div}\,u + \gamma \, w = \bar H,
\end{equation}
where $\nabla \bar H = \bar F$. Combining the last equation with (\ref{system_lin})$_2$ we arrive at
\begin{equation} \label{trans}
\bar \gamma w + w_{x_1} + (\bar u+u_0) \nabla w = H,
\end{equation}
where $\bar \gamma = \frac{\gamma}{\nu + 2 \mu}$ and
\begin{equation} \label{H}
H = \frac{\bar H}{\nu + 2\mu}+G.
\end{equation}
The equation (\ref{trans}) makes it possible to estimate the $W^1_p$-norm of the density
in terms of $W^1_p$ - norm of $H$.
The latter will be controlled since (\ref{nablaH}) enables us to bound $||\nabla H||_{L_p}$
and $||H||_{L_p}$ using interpolation and the energy estimate (\ref{ene1}).
The details are presented in the proof of lemma \ref{lemH}, but first we estimate
$||w||_{W^1_p}$ in terms of $H$. The result is stated in the following lemma
\begin{lem}
Assume that $w$ satisfies the equation (\ref{trans}) with $H \in W^1_p$. Then
\begin{equation} \label{w_w1p}
||w||_{W^1_p} \leq C \, \big[ ||H||_{W^1_p} + ||w_{in}||_{W^1_p(\Gamma_{in})} \big].
\end{equation}
\end{lem}
\textbf{Proof.} In order to find a bound on $||w||_{L_p}$ we multiply (\ref{trans}) by $|w|^{p-2} w$
and integrate over $\Omega$. Integrating by parts
and next using the boundary conditions we get
$$
\int_{\Omega} |w|^{p-2} w \,w_{x_1} \,dx = \frac{1}{p} \int_{\Omega} \partial_{x_1} |w|^p \,dx
= \frac{1}{p} \int_{\Gamma_{out}} |w|^p \, d\sigma - \frac{1}{p} \int_{\Gamma_{in}} |w|^p \, d\sigma ,
$$
since $n^{(1)} \equiv 0$ on $\Gamma_0$, $n^{(1)} \equiv -1$ on $\Gamma_{in}$
and $n^{(1)} \equiv 1$ on $\Gamma_{out}$. Similarily, applying the boundary conditions we get
$$
\int_{\Omega} (\bar u+u_0) \cdot (|w|^{p-2} w \nabla w) \,dx =
\frac{1}{p} \int_{\Omega} (\bar u+u_0) \cdot \nabla |w|^p \,dx =
$$$$
- \frac{1}{p} \int_{\Omega} {\rm div}\, (\bar u+u_0) \, |w|^p \,dx
+ \frac{1}{p} \int_{\Gamma_{out}} u_0^{(1)} \, |w|^p \, d\sigma - \frac{1}{p} \int_{\Gamma_{in}} u_0^{(1)} \, |w|^p \, d\sigma.
$$
Thus multiplying (\ref{trans}) by $|w|^{p-2} w$ we get
\begin{equation} \label{w1}
\begin{array}{c}
\bar \gamma ||w||_{L_p}^p - \frac{1}{p} \int_{\Omega} {\rm div}\, (\bar u+u_0) \, |w|^p \,dx + \frac{1}{p} \int_{\Gamma_{out}} |w|^p \, (1+u_0^{(1)})  \,d\sigma \leq \\
\leq ||H||_{L_p} \, ||w||_{L_p}^{p-1} + \frac{1}{p} \int_{\Gamma_{in}} |w_{in}|^p \, (1+u_0^{(1)}) \,d\sigma.
\end{array}
\end{equation}
By the imbedding theorem the smallness of $||\bar u+u_0||_{W^2_p}$ implies $1+u_0^{(1)}>0$ a.e. in $\Omega$
and $\bar \gamma - ||{\rm div}\, (u+u_0)||_{\infty} > 0$.
Thus the boundary term on the l.h.s. is positive and the term with ${\rm div}\, (u+u_0)$ can be combined
with the first term of the l.h.s, what yields
$$
C \, ||w||_{L_p}^p \leq
||H||_{L_p} \, ||w||_{L_p}^{p-1} + C \, ||w_{in}||_{L_p(\Gamma_{in})}^p,
$$
and so
\begin{equation} \label{w}
||w||_{L_p} \leq C \, \big[ ||H||_{L_p} + ||w_{in}||_{L_p(\Gamma_{in})} \big] .
\end{equation}
The derivatives of the density are estimated in a similar way. In order to find a bound on $w_{x_i}$
we differentiate (\ref{trans}) with respect to $x_i$. If we assume that $w \in W^1_p$ then (\ref{trans})
implies $\tilde u \cdot \nabla w \in W^1_p$, where
\begin{equation} \label{tilde_u}
\tilde u := [1 + (\bar u + u_0)^{(1)}, (\bar u + u_0)^{(2)},(\bar u + u_0)^{(3)}].
\end{equation}
Thus $\tilde u \cdot \nabla w_{x_i} := (\tilde u \cdot \nabla w)_{x_i} - \tilde u_{x_i} \cdot \nabla w \in L_p$.
Hence we can differentiate (\ref{trans}) with respect to $x_i$, multiply by $|w_{x_i}|^{p-2} w_{x_i}$ and integrate.
Since $\tilde u_{x_i} = (\bar u+u_0)_{x_i}$, we have
$$
\int_{\Omega} \tilde u_{x_i} \cdot (|w_{x_i}|^{p-2} w_{x_i} \nabla w) \,dx \leq
||\nabla (\bar u+u_0)||_{L_{\infty}} \, ||\nabla w||_{L_p}^p \leq C \, ||\bar u+u_0||_{W^2_p} \, ||\nabla w||_{L_p}.
$$
Next, since $\tilde u \cdot \nabla w_{x_i} \in L_p$, we can write   
$$
\int_{\Omega} \tilde u \cdot |w_{x_i}|^{p-2} w_{x_i} \nabla w_{x_i} \,dx =
\frac{1}{p} \int_{\Omega} \tilde u \cdot \nabla |w_{x_i}|^p \,dx 
= - \frac{1}{p} \int_{\Omega} |w_{x_i}|^p \, {\rm div}\, \tilde u \,dx + \frac{1}{p} \int_{\Gamma} |w_{x_i}|^p \, \tilde u \cdot n \, d\sigma= 
$$$$
= - \frac{1}{p} \int_{\Omega} |w_{x_i}|^p \, {\rm div}\, \tilde u \,dx - \frac{1}{p} \int_{\Gamma_{in}} |w_{in,x_i}|^p \, (1+u_0^{(1)}) \,d\sigma
+ \frac{1}{p} \int_{\Gamma_{out}} |w_{x_i}|^p \, (1+u_0^{(1)}) \,d\sigma.
$$
For $i=2,3$ we have $w_{in,x_i} \in L_p(\Gamma_{in})$ and hence the above defines the trace of $|w_{x_i}|^p$ on $\Gamma_{out}$.
We arrive at
\begin{equation}  \label{wxi}
\begin{array}{c}
\bar \gamma ||w_{x_i}||_{L_p}^p - \frac{1}{p} \int_{\Omega} {\rm div} \, (\bar u+u_0) \, |w_{x_i}|^p \,dx + \frac{1}{p} \int_{\Gamma_{out}} |w_{x_i}|^p \, (1+u_0^{(1)})  \,d\sigma \leq \\
\leq ||H_{x_i}||_{L_p} \, ||w_{x_i}||_{L_p}^{p-1} + \frac{1}{p} \int_{\Gamma_{in}} |w_{in,x_i}|^p \, (1+u_0^{(1)}) \,d\sigma + C \, ||\bar u+u_0||_{W^2_p} ||\nabla w||_{L_p}^p.
\end{array}
\end{equation}
For $i=2,3$ it gives directly the bound on $||w_{x_i}||_{L_p}$.
In order to estimate $w_{x_1}$ we start the same way differentiating (\ref{trans}) with respect to $x_1$
and multiplying by $|w_{x_1}|^{p-2} w_{x_1}$. The difference in comparison to $w_{x_2}$ and $w_{x_3}$ is that
$w_{x_1}$ is not given on $\Gamma_{in}$.
In order to overcome this difficulty we can observe that
on $\Gamma_{in}$ the equation (\ref{trans}) reduces to
$$
\bar \gamma w_{in} + (\bar u+u_0)^{(2)} \, w_{in,x_2} + (\bar u+u_0)^{(3)} \, w_{in,x_3} + [1 + (\bar u+u_0)^{(1)}] \,w_{x_1} = H,
$$
what can be rewritten as
$$
w_{x_1} = \frac{1}{1+(\bar u+u_0)^{(1)}} \, \big[ H - \bar \gamma w_{in} - (\bar u+u_0)_{\tau} \cdot \nabla_{\tau} w_{in} \big].
$$
Thus we have
$$
||w_{x_1}||_{L_p(\Gamma_{in})} \leq C \, \big[ ||H|_{\Gamma_{in}}||_{L_p(\Gamma_{in})} + ||w_{in}||_{W^1_p(\Gamma_{in})} \big].
$$
Using this bound in (\ref{wxi}), $i=1$, we arrive at the estimate
\begin{equation} \label{wx1}
||w_{x_1}||_{L_p}^p \leq C \, \big[ ||H_{x_1}||_{L_p} \, ||w_{x_1}||_{L_p}^{p-1} + ||\bar u+u_0||_{W^2_p} \, ||\nabla w||_{L_p}^p
+ ||H||_{L_p(\Gamma_{in})}^p + ||w_{in}||_{W^1_p(\Gamma_{in})}^p \big].
\end{equation}
The boundary term $||H||_{L_p(\Gamma_{in})}$ can by replaced by $||H||_{W^1_p}$ due to the trace theorem.
Thus combining (\ref{wxi}) (for $x_2$ and $x_3$) with (\ref{wx1}) we get
\begin{equation}
||\nabla w||_{L_p}^p \leq C \, \big[ ||\nabla H||_{L_p} ||\nabla w||_{L_p}^{p-1}
+ ||\bar u+u_0||_{W^2_p} ||\nabla w||_{L_p}^p + ||H||_{W^1_p}^p + ||w_{in}||_{W^1_p(\Gamma_{in})}^p \big].
\end{equation}
The term $||u+u_0||_{W^2_p} ||\nabla w||_{L_p}^p$ can be put on the l.h.s. due to the smallness assumption
and thus we get
\begin{equation}
||\nabla w||_{L_p} \leq C \, [ ||H||_{W^1_p} + ||w_{in}||_{W^1_p(\Gamma_{in})} ],
\end{equation}
what combined with (\ref{w}) yields
\begin{equation} \label{w_w1p_1}
||w||_{W^1_p} \leq C \, \big[ ||H||_{W^1_p} + ||H||_{L_p(\Gamma_{in})} + ||w_{in}||_{W^1_p(\Gamma_{in})} \big].
\end{equation}
Applying again the trace theorem to the term $||H||_{L_p(\Gamma_{in})}$
we arrive at (\ref{w_w1p}). $\square$

The next step is to estimate $H$ in terms of the data. The result is in the following
\begin{lem}   \label{lemH}
Let $H$ be defined in (\ref{H}). Then $\forall \delta>0$ we have
\begin{equation}    \label{lemH_teza}
||H||_{W^1_p} \leq
\delta ||u||_{W^2_p} + C(\delta) [||F||_{L_p} + ||G||_{W^1_p} + ||B||_{W^{1-1/p}_p(\Gamma)} + ||w_{in}||_{W^1_p(\Gamma_{in})} ].
\end{equation}
\end{lem}
\textbf{Proof.}
Applying first the interpolation inequality (\ref{int1}) and then the estimate (\ref{ene1})
we get
\begin{equation}     \label{Hlp}
||H||_{L_p} \leq \delta_1 ||\nabla H||_{L_p}
+ C(\delta_1) \, [ ||F||_{L_2} + ||G||_{L_2} + ||B||_{L_2(\Gamma)} ].
\end{equation}
Next, by (\ref{nablaH}) we have
$$
||\nabla H||_{L_p} \leq C \, [ ||F||_{L_p} + ||G||_{W^1_p} + ||A||_{W^2_p} + ||\partial_{x_1} \phi||_{L_p} ],
$$
where $u = \nabla \phi + A$ is the Helmholtz decomposition.
Now we use the bound (\ref{Aw2p}) on $||A||_{W^2_p}$. We obtain a term
$||u||_{W^{1-1/p}_p(\Gamma)}$, that we estimate using the trace theorem and the interpolation
inequality (\ref{int1}). The same inequality is applied to estimate $||\partial_{x_1}\phi||_{L_p}$.
We arrive at
\begin{equation}    \label{nablahlp}
\begin{array}{c}
||\nabla H||_{L_p} \leq C \, [ ||F||_{L_p} + ||G||_{W^1_p} + ||B||_{W^{1-1/p}_p(\Gamma)} + ||w_{in}||_{W^1_p(\Gamma_{in})} ] \\
+ \delta_1 ||u||_{W^2_p} + C(\delta_1) [||F||_{L_2} + ||G||_{L_2} + ||B||_{L_p(\Gamma)} ].
\end{array}
\end{equation}
Combining (\ref{Hlp}) and (\ref{nablahlp}) we get (\ref{lemH_teza}) $\square.$

Now we are ready to show the \emph{a priori} estimate in $W^2_p \times W^1_p$
on the solution of the linear problem.

\begin{lem}
Let $(u,w)$ be a solution to (\ref{system_lin}) with 
$(F,G,B,w_{in},\bar u) \in L_p \times W^1_p \times W^{1-1/p}_p(\Gamma) \times W^1_p(\Gamma_{in}) \times W^2_p$,
with $||\bar u||_{W^2_p}$ small enough and $f$ large enough. Then
\begin{equation} \label{est_lin_w2p}
||u||_{W^2_p} + ||w||_{W^1_p} \leq C \, [||F||_{L_p} + ||G||_{W^1_p} + ||B||_{W^{1-1/p}_p(\Gamma)} + ||w_{in}||_{W^1_p(\Gamma_{in})}].
\end{equation}
\end{lem}
\textbf{Proof.} If $(u,w)$ is a solution to (\ref{system_lin}), then in particular the velocity
satisfies the Lame system
\begin{eqnarray}
\begin{array}{lcr}
\partial_{x_1} u -\mu \Delta  u - (\nu + \mu) \nabla {\rm div}\, u =  F - \gamma \nabla  w & \mbox{in} & \Omega,\\
n\cdot 2\mu {\bf D}( u)\cdot \tau_i +f \ u \cdot \tau_i = B_i, \quad i=1,2
&\mbox{on} & \Gamma, \\
n\cdot  u = 0 & \mbox{on} & \Gamma.\\
\end{array}
\end{eqnarray}
The classical theory of elliptic equations (Agmon,Douglis,Nirenberg \cite{ADN1},\cite{ADN2}) yields
$$
||u||_{W^2_p} \leq C \, [||F||_{L_p} + ||w||_{W^1_p} + ||B||_{W^{1-1/p}_p} + ||u||_{W^1_p}].
$$
Applying the interpolation inequality (\ref{int1}) to the term $||u||_{W^1_p}$ and then the energy estimate (\ref{ene1})
we get
\begin{equation} \label{est_lin_w2p_1}
||u||_{W^2_p} \leq C \, [||F||_{L_p} + ||G||_{W^1_p} + ||w||_{W^1_p} + ||B||_{W^{1-1/p}_p} + ||w_{in}||_{L_2(\Gamma_{in})}].
\end{equation}
In order to complete the proof we combine (\ref{w_w1p}) and (\ref{lemH_teza}) obtaining
\begin{equation}
||w||_{W^1_p} \leq \delta ||u||_{W^2_p} + C(\delta) [||F||_{L_p} + ||G||_{W^1_p} + ||B||_{W^{1-1/p}_p(\Gamma)} + ||w_{in}||_{W^1_p(\Gamma_{in})} ],
\end{equation}
and choosing for example $\delta = \frac{1}{2C}$ where $C$ is the constant from (\ref{est_lin_w2p_1})
we arrive at (\ref{est_lin_w2p}). $\square$
\section{Solution of the linear system}  \label{sec_sol_lin}
In this section we show the existence of the sequence $(u^n,w^n)$ defined in (\ref{system_seq}).
To this end we have to solve the linear system (\ref{system_lin}) where
$(F,G,\bar u,u_0) \in L_p \times W^1_p \times W^2_p \times W^2_p$ are given functions
such that $\bar u \cdot n = 0$ on $\Gamma$. First we apply the Galerkin method to prove
the existence of a weak solution and next we show that this solution is strong.
For simplicity we will denote $\bar u + u_0$ by $\bar u$.
\subsection{Weak solution}
Let us recall the definition of the space $V$ (\ref{def_V}).
A natural definition of a weak solution to the system (\ref{system_lin}) is a couple
$(u,w) \in V \times L_\infty(L_2)$ such that
\begin{eqnarray} \label{weak1}
\int_{\Omega} \{ v \cdot \partial_{x_1} u + 2 \mu {\bf D}(u) : \nabla \,v + \nu \, {\rm div}\,u \, {\rm div}\,v
- \gamma w \, {\rm div}\,v \} \,dx
+ \int_{\Gamma} f (u \cdot \tau_i) \, (v \cdot \tau_i) \,d\sigma = \nonumber\\
= \int_{\Omega} F \cdot v \,dx + \int_{\Gamma} B_i (v \cdot \tau_i) \,d\sigma
\end{eqnarray}
is satisfied $\forall \; v \in V$ and (\ref{system_lin})$_2$ is satisfied in ${\cal D'}(\Omega)$, i.e.
$\forall \; \phi \in \bar C^{\infty}(\Omega)$:
\begin{equation} \label{weak2}
-\int_{\Omega} w \tilde u \cdot \nabla \phi \,dx - \int_{\Omega} w \phi \, {\rm div}\, \tilde u \,dx
+ \int_{\Gamma_{out}} w \, \phi \,d\sigma =
\int_{\Omega} \phi (G - {\rm div}\, u) \,dx + \int_{\Gamma_{in}} w_{in} \phi \, d\sigma,
\end{equation}
where $\tilde u$ is defined in (\ref{tilde_u}).
Let us introduce an orthonormal basis of V: $\{\omega_i\}_{i=1}^{\infty}$. We consider finite dimensional spaces:
\mbox{$V^N = \{ \sum_{i=1}^N \alpha_i \omega_i: \; \alpha_i \in \mathbf{R} \} \subset V$}.
The sequence of
approximations to the velocity will be searched for in a standard way as
$
u^N = \sum_{i=1}^N c_i^N \, \omega_i.
$
Due to the equation (\ref{system_lin})$_2$ we have to define the approximations to the density
in an appropriate way. Namely, we set $w^N = S(G^N - {\rm div}\, u^N)$,
where $S:L_2(\Omega) \to L_{\infty}(L_2)$ is defined as
\begin{equation} \label{def_S}
w=S(v) \iff \left\{ \begin{array}{lcr}
\partial_{x_1}w + \bar u \cdot \nabla  w = v & \textrm{in} & {\cal D'}(\Omega), \\
w = w_{in} & \textrm{on} & \Gamma_{in}.
\end{array} \right.
\end{equation}
We want the image of $S$ to be in the space $L_{\infty}(L_2)$
so that we can apply the theory of transport equation treating $x_1$ as a 'time' variable
to show that $S$ is well defined.
In order to solve the system on the r.h.s. of (\ref{def_S}) we can search for a change of variables
$x = \psi(z)$ satisfying the identity
\begin{equation} \label{change_id}
\partial_{z_1} = \partial_{x_1} + \bar u \cdot \nabla_x.
\end{equation}
We construct the mapping $\psi$ in the following
\begin{lem}  \label{lem_change}
Let $||\bar u||_{W^2_p}$ be small enough. Then there exists a set $U \subset \mathbb{R}^3$
and a diffeomorphism $x=\psi(z)$ defined on $U$ such that $\Omega = \psi(U)$ and (\ref{change_id}) holds.
Moreover, if $z_n \to z$ and $\psi(z_n) \to \Gamma_0$ then $n^1(z) = 0$, where $n$ is the outward normal
to $U$.

\end{lem}
Before we start with the proof we shall make one remark.
The last condition states that the first component of the normal to $\psi^{-1}(\Gamma_0)$ vanishes,
but since $\psi$ is defined only on $U$ we formulate this condition using the limits.
It means simply that the image $U = \psi^{-1}(\Omega)$ is also a cylinder with a flat wall. It will be important in the construction
of the operator $S$.

\textbf{Proof of lemma \ref{lem_change}.}
The identity (\ref{change_id}) means that $\psi$ must satisfy
\begin{equation} \label{psi_z1}
\frac{\partial \psi^{1}}{\partial z_1} = 1 + \bar u^1 (\psi), \quad
\frac{\partial \psi^{2}}{\partial z_1} = \bar u^2 (\psi), \quad
\frac{\partial \psi^{3}}{\partial z_1} = \bar u^3 (\psi).
\end{equation}
A natural condition is that $\psi(\Gamma_{in}) = \Gamma_{in}$.
Thus we can search for $\psi(z_1,z_2,z_3) = \psi_{z_2,z_3}(z_1)$,
where for all $(z_2,z_3)$ such that $(z_2,z_3,0) \in \Gamma_{in}$
the function $\psi_{z_2,z_3}(\cdot)$ is a solution to a system of ODE:
\begin{equation}  \label{ode}
\left\{ \begin{array}{l}
\partial_s \psi_{z_2,z_3}^1 = 1 + \bar u^1(\psi_{z_2,z_3}), \quad
\partial_s \psi_{z_2,z_3}^2 = \bar u^2(\psi_{z_2,z_3}), \quad
\partial_s \psi_{z_2,z_3}^3 = \bar u^3(\psi_{z_2,z_3}),\\
\psi_{z_2,z_3}(0) = (0,z_2,z_3).
\end{array} \right.
\end{equation}
The r.h.s of the system (\ref{ode}) is a Lipschitz function with a constant
$K = ||\nabla \bar u||_{\infty}$ and thus provided that $||\bar u||_{W^2_p}$ is small enough
the system (\ref{ode}) has a unique solution defined on some interval $(0,b_{z_1,z_2})$,
where $b_{z_1,z_2}$ depends on $z_2,z_3$ and $||\nabla \bar u||_{\infty}$. Provided that the latter
is small enough the function $\psi(z) = \psi_{z_2,z_3}(z_1)$ will be defined on $U$ such that
$\Omega = \psi(U)$.

Now we show that $\psi(z) = \psi_{z_2,z_3}(z_1)$ is a diffeomorphizm. The derivatives with
respect to $z_1$ are given by (\ref{psi_z1}) and the remaining derivatives can be expressed in terms
of $\bar u$ so we can see that $J \, \psi  = 1 + E(\bar u)$, where $E(\bar u)$ is small
(and thus $J \, \psi > 0$) provided that $||\bar u||_{W^2_p}$ is small.

To see that $\psi$ is $1-1$ we can write it in a form  $\psi(z)=z+\epsilon(z)$,
where $||\nabla \epsilon||_{L_\infty}$ is small.
Assume that $\psi(z^1) = \psi(z^2)$ and $z^1 \neq z^2$. Then there exists $i$ such that
$|z^1_i - z^2_i| \geq \frac{1}{3} |z^1-z^2|$ (the lowercase denotes the coordinate). On the other hand, we have
$|z^1_i - z^2_i| = |\epsilon_i(z^1) - \epsilon_i(z^2)| \leq ||\nabla \epsilon||_{L_{\infty}} \, |z^1-z^2|$,
what contradicts the smallness of $||\nabla \epsilon||_{L_\infty}$.

We have shown that the mapping $\psi$ given by (\ref{ode}) is a diffeomorphizm
defined on $U$ such that $\psi(U) = \Omega$. Let us denote $\phi = \psi^{-1}$.
Now it is natural to define the
subsets of $\partial_U$ as $\partial_U = U_{in} \cup U_{out} \cup U_0$ where $U_{in} = \Gamma_{in}$,
$U_{out} = \{z: \; z = {\rm lim} \, \phi(x_n), \; x_n \to \Gamma_{out}\}$ and
$U_{0} = \{z: \; z = {\rm lim} \, \phi(x_n), \; x_n \to \Gamma_{0}\}$.

In order to complete the proof we have to show that
$n^1(z)=0$ for $z \in U_0$. But to this end it is enough to observe that
$$
D \psi(z) ([1,0,0]) = [1+\bar u^1(x),\bar u^2(x),\bar u^3(x)],
$$
where $x=\psi(z)$. But for $x \in \Gamma_0$ the vector on the r.h.s is tangent to $\Gamma_0$
since $\bar u \cdot n =0$.
We can conclude that on $U_0$ the image in $\psi$ of a straight line
$\{(s,z_2,z_3): \, s \in (0,b)\}$ is a curve tangent to $\Gamma_0$, and thus
$U_0$ is a sum of such lines and so we have $n^1(z)=0$. The proof
of lemma \ref{lem_change} is completed. $\square.$

Now we can define $S(v)$ for a continuous function $v$ as
\begin{equation} \label{S}
S(v)(x) = w_{in}(0,\phi_2(x),\phi_3(x))
+ \int_{0}^{\phi_1(x)} v (\psi(s,\phi_2(x),\phi_3(x))) \,ds.
\end{equation}
The condition $n^1=0$ on $\phi(\Gamma_0)$ guarantees that a straight line
$(s,z_1,z_2): s \in (0,b)$ has a picture in $\Omega$ and thus we integrate along a curve contained in $\Omega$.
It means that $S$ is well defined for continuous functions defined on $\Omega$ and
the construction of $\psi$ clearly ensures that $S$ satisfies (\ref{def_S}). Next we have to extend $S$
on $L_2(\Omega)$. To this end we need an estimate in $L_{\infty}(L_2)$.
It is given by the following
\begin{lem}  \label{lem_est_S}
Let $S$ be defined in (\ref{S}). Then
\begin{equation} \label{est_S}
||S(v)||_{L_{\infty}(L_2)} \leq C \, [||w_{in}||_{L_2(\Gamma_{in})} + ||v||_{L_2(\Omega)}].
\end{equation}
\end{lem}
\textbf{Proof}.
Let $\Omega_{x_1}$ denote an $x_1$ - cut of $\Omega$ and let $\bar x := (x_2,x_3)$.
Then by (\ref{S}) we have
$$
||S(v)||_{L_2(\Omega_{x_1})}^2 =
\int_{\Omega_{x_1}} \big[ w_{in}(0,\phi_2(x),\phi_3(x)) + \int_0^{\phi_1(x)} v(\psi(s,\phi_2(x),\phi_3(x))) \,ds \big]^2 \, d \bar x
$$$$
\leq 2 ||w_{in}||_{L_2(\Gamma_{in})}^2 + C \int_{\Omega_{x_1}} \int_0^{\phi_1(x)} v^2(\psi(s,\phi_2(x),\phi_3(x))) \,ds \,d\bar x
\leq C \, [ ||w_{in}||_{L_2(\Gamma_{in})}^2 + ||v||_{L_2(\Omega)}^2 ].
$$
The above holds for every $x_1 \in (0,L)$ what implies (\ref{est_S}). $\square$

Now we can define $S(v)$ for $v \in L_2(\Omega)$ using a standard density argument.
Let us take a sequence of smooth functions $v_n \to v$ in $L_2(\Omega)$. By (\ref{est_S}) the sequence
$S(v_n)$ satisfies
\begin{equation} \label{est_svn}
||S(v_n)||_{L_{\infty}(L_2)} \leq C \, [ ||w_{in}||_{L_2(\Gamma_{in})} + {\rm sup}_n ||v_n||_{L_2} ].
\end{equation}
The bound on the r.h.s. is uniform in $n$ and thus $S(v_n) \rightharpoonup^* \eta$
in $L_{\infty}(L_2)$, and $\eta$ satisfies the estimate (\ref{est_S}).
In particular for $\phi \in \bar C^{\infty}(\Omega)$ we have
$$
\int_{\Omega} S(v_n) \tilde u \cdot \nabla \phi \,dx \to \int_{\Omega} \eta \tilde u \cdot \nabla \phi \,dx
\quad
{\rm and}
\quad
\int_{\Omega} S(v_n) \phi \, {\rm div}\, \tilde u \,dx \to \int_{\Omega} \eta \phi \, {\rm div}\, \tilde u \,dx.
$$
In order to show that $\eta = S(v)$, i.e. $\eta$ solves the system on the r.h.s. of (\ref{def_S})
we have to show that $\int_{\Gamma_{out}} S(v_n) \, \phi \,d\sigma \to \int_{\Gamma_{out}} \eta \, \phi \, d\sigma$.
To this end notice that the proof of lemma \ref{lem_est_S} implies in particular that
$||S(v_n)||_{L_2(\Gamma_{out})}$ satisfies the estimate (\ref{est_svn}). Thus
$S(v_n) \rightharpoonup \zeta$ in $L_2(\Gamma_{out})$ for some $\zeta \in L_2(\Gamma_{out})$,
and in particular $\int_{\Gamma_{out}}S(v_n) \phi \,d\sigma \to \int_{\Gamma_{out}} \zeta \phi \,d\sigma$.
We have to verify that $\eta|_{\Gamma_{out}} = \zeta$. This would not be obvious if we only had
$S(v_n) \in L_{\infty}(L_2)$, but indeed the proof of lemma $\ref{lem_est_S}$
implies a stronger condition that supremum (not only the essential supremum) of $||S(v_n)||_{L_2(\Omega_{x_1})}$
is bounded, thus we must have $\zeta = \eta|_{\Gamma_{out}}$.
We have shown that $\tilde u \cdot \nabla \eta = v$ in ${\cal D'}(\Omega)$, thus indeed $\eta = S(v)$.

Having the operator $S$ well defined we are ready to proceed with the Galerkin method.
Taking $F = F^N$, $u = u^N = \sum_i c_i^N \, \omega_i$, $v = \omega_k, \quad k=1 \ldots N$ and $w = w^N = S(G^N - {\rm div}\, u^N)$ in (\ref{weak1}),
where $F^N$ and $G^N$ are orthogonal projections of $F$ and $G$ on $V^N$,
we arrive at a system of $N$ equations
\begin{equation}  \label{system_aprox}
B^N(u^N,\omega_k) = 0, \quad k = 1 \ldots N,
\end{equation}
where $B^N:V^N \to V^N$ is defined as
\begin{equation}
\begin{array}{c}
B^N(\xi^N,v^N) = \int_{\Omega} \big\{ \xi^N \partial_{x_1}v^N + 2\mu\mathbf{D}(\xi^N) : \nabla v^N + {\rm div}\,\xi^N \, {\rm div}\,v^N \big\}\,dx \\
- \gamma \int_{\Omega} S(G^N-{\rm div}\,\xi^N) \, {\rm div}\,v^N \,dx 
+ \int_{\Gamma} [f \, (\xi^N \cdot \tau_j) - B_i] \, (v^N \cdot \tau_j) \,d\sigma - \int_{\Omega} F^N \cdot v^N \,dx.
\end{array}
\end{equation}
Now, if $u^N$ satisfies (\ref{system_aprox}) for $k = 1 \ldots N$ and $w^N = S(G^N-{\rm div}\,u^N)$, then a pair
$(u^N,w^N)$ satisfies (\ref{weak1}) - (\ref{weak2}) for $(v,\phi) \in (V^N \times \bar C^{\infty}(\Omega))$.
We will call such a pair an approximate solution to (\ref{weak1}) - (\ref{weak2}).

The following lemma gives existence of solution to the system (\ref{system_aprox}):
\begin{lem}
Let $F,G \in L^2(\Omega)$, $w_{in} \in L_2(\Gamma_{in})$, $B \in L_2(\Gamma)$ and assume that $f$ is large enough and $||\bar u||_{W^2_p}$
is small enough. Then there exists $u^N \in V^N$ satisfying (\ref{system_aprox}) for $k=1 \ldots N$.
Moreover,
\begin{equation}  \label{est_uN}
||u^N||_{H^1} \leq C(DATA).
\end{equation}
\end{lem}
\textbf{Proof.}
In order to solve the system (\ref{system_aprox}) we will apply a well-known result in finite-dimensional
Hilbert spaces, lemma \ref{lem_P} in the Appendix. Thus we define the operator
$P^N:V^N \to V^N$ as
\begin{equation}   \label{P}
P^N(\xi^N) = \sum_k B^N(\xi^N,\omega_k) \omega_k \quad \textrm{for} \quad \xi^N \in V^N.
\end{equation}
In order to apply lemma \ref{lem_P} we have to show that $\big(P(\xi^N),\xi^N\big) > 0$ on some sphere in $V^N$.
Since $B^N(\cdot,\cdot)$ is linear with respect to the second variable, we clearly have
\begin{equation}
\begin{array}{c}
\big( P(\xi^N),\xi^N \big) = B^N(\xi^N,\xi^N) =
\underbrace{2\mu \int_{\Omega} D^2(\xi^N) \,dx + \nu \int_{\Omega} {\rm div}^2 \xi^N \,dx}_{I_1} \\
+ \underbrace{\int_{\Omega} \xi^N \partial_{x_1}\xi^N \,dx + \int_{\Gamma} f (\xi^N \cdot \tau_i)^2 \,d\sigma}_{I_2}
\underbrace{-\gamma \int_{\Omega} S(G^N - {\rm div}\, \xi^N) \, {\rm div}\,\xi^N  \,dx}_{I_3} \\
- \int_{\Omega} F \cdot \xi^N \,dx - \int_{\Gamma} B_i \, (\xi^N \cdot \tau_i) \,d\sigma .
\end{array}
\end{equation}
Using the Korn inequality similarily as in the proof of the energy estimate (\ref{ene1}) we get
\begin{equation} \label{est_P_1}
I_1+I_2 \geq C \, ||\xi^N||_{H^1}^2
\end{equation}
for $f$ large enough. We have to find a bound on $I_3$.
Denoting $\eta^N = S (G^N - {\rm div}\, \xi^N)$ we have
\begin{equation} \label{I3_1}
-\int_{\Omega} \eta^N \, {\rm div}\,\xi^N \,dx =
\int_{\Omega} \eta^N ( \partial_{x_1} \eta^N + \bar u \cdot \nabla \eta^N )\,dx
-\int_{\Omega} \eta^N \, G^N \,dx.
\end{equation}
Using (\ref{est_S}) we get
\begin{equation}
-\int_{\Omega} \eta^N \, G^N \,dx \geq - ||\eta^N||_{L^2} \, ||G^N||_{L^2}
\geq - C \, ||G^N||_{L_2} \, (||G^N||_{L_2} + ||\xi^N||_{H^1} + ||w_{in}||_{L_2(\Gamma_{in})}).
\end{equation}
With the first integral on the r.h.s of (\ref{I3_1}) we have
\begin{equation} \label{I3_2}
\begin{array}{c}
\int_{\Omega} \eta^N ( \partial_{x_1} \eta^N + \bar u \cdot \nabla \eta^N )\,dx
= \int_{U} \eta^N(z) \partial_{z_1} \eta^N(z) J \psi(z) \,dz = \\
= \int_U \eta^N(z) \partial_{z_1} \eta^N(z) \,dz +
\int_U \eta^N(z) \partial_{z_1} \eta^N(z) [J \psi(z) -1] \,dz.
\end{array}
\end{equation}
The first integral can be rewritten as a boundary integral and since $n^1(z)=0$ on $\phi(\Gamma_0)$,
it reduces to
$$
\frac{1}{2} \int_{\partial U} [\eta^N(z)]^2 n^1(z) d\sigma(z) =
-\frac{1}{2} \int_{U_{in}} [\eta^N(z)]^2 \, d\sigma(z)
+ \frac{1}{2} \int_{U_{out}} [\eta^N(z)]^2 \, d\sigma(z) \geq
- \int_{\Gamma_{in}} w_{in}^2 \,d\sigma(x).
$$
In the last passage we used the fact that $\phi|_{\Gamma_{in}}$ is the identity and that
$n^1(z)>0$ on $U_{out}$, what is true provided that $\phi$ does not differ too much
from the identity on $\Gamma_{out}$, what in turn holds under the smallness assumptions on $\bar u$.

With the second integral on the r.h.s. of (\ref{I3_2}) we have
$$
\int_U \eta^N(z) \partial_{z_1} \eta^N(z) [J \psi(z) -1] \,dz \geq
- {\rm sup}_U |J\psi-1| \, \int_U \eta^N(z) (G^N - {\rm div}_x \,\xi^N)(z) \,dz \geq
$$$$
\geq - E \, ||\eta^N||_{L_2(U)} \, [ ||G^N||_{L_2(U)} + ||{\rm div}_x \xi^N||_{L_2(U)} ] \geq
- E \, [||G^N||_{L_2(\Omega)}^2 + ||\xi^N||_{H^1(\Omega)}^2 + ||w_{in}||_{L_2(\Gamma_{in})}^2 ].
$$
Combining this estimate with (\ref{est_P_1}) we get
\begin{equation}
\big( P^N(\xi^N),\xi^N \big) \geq C \,\big[ ||\xi^N||_{H^1(\Omega)}^2 - D \, ||\xi^N||_{H^1(\Omega)} - D^2 \big],
\end{equation}
where $D = ||F||_{L^2(\Omega)}+||G||_{L^2(\Omega)} + ||w_{in}||_{L_2(\Gamma_{in})} + ||B||_{L_2(\Gamma)} $.
Thus there exists $\tilde C = \tilde C(\mu,\Omega,D)$ such that
$\big( P^N(\xi^N),\xi^N \big) > 0 \quad \textrm{for} \quad ||\xi||=\tilde C$,
and applying lemma \ref{lem_P} we conclude that
$\exists \xi^*: \quad P^N(\xi^*)=0 \quad \textrm{and} \quad ||\xi^*|| \leq \tilde C$.
Moreover, since $\{\omega_k\}_{k=1}^N$ is the basis of $V^N$, we have
$P^N(\xi^*)=0 \iff (B^N\xi^*,\omega_k)=0, \quad k=1 \ldots N$. Thus $\xi^*$ is a solution to
(\ref{system_aprox}).
$\square$.

Now showing the existence of the weak solution is straightforward. The result is in the following
\begin{lem} \label{lem_weak}
Assume that $F,G \in L_2(\Omega)$, $w_{in} \in L_2(\Gamma_{in})$, $B \in L_2(\Gamma)$. 
Assume further that $f$ is large enough and $||\bar u||_{W^2_p}$ is small enough.
Then there exists $(u,w) \in V \times W$ that is a weak solution to the system (\ref{system_lin}).
Moreover, the weak solution satisfies the estimate (\ref{ene1}).
\end{lem}
\textbf{Proof.} The estimates (\ref{est_S}) and (\ref{est_uN}) imply that
$||u^N||_{H^1}+||w^N||_{L_{\infty}(L_2)} \leq C(DATA)$. Thus
$$
u^N \rightharpoonup u \; {\rm in} \; H^1 \quad {\rm and} \quad w^N \rightharpoonup^* w \; {\rm in} \; L_{\infty}(L_2)
$$
for some $(u,w) \in H^1 \times L_{\infty}(L_2)$.
It is very easy to verify that $(u,w)$ is a weak solution.
First, passing to the limit in (\ref{weak1}) for $(u^N,w^N)$ we see that
$u$ satisfies (\ref{weak1}) with $w$. On the other hand, taking the limit in (\ref{weak2})
we verify that $w = S(G - {\rm div}\, u)$.
We conclude that $(u,w)$ satisfies (\ref{weak1}) - (\ref{weak2}), thus we have the weak solution.
To show the boundary condition on the density we can rewrite the r.h.s of (\ref{def_S}) as
\begin{equation} \label{def_S_1}
\left\{ \begin{array}{lcr}
w_{x_1} + \frac{\bar u^{(2)}}{1+\bar u^{(1)}} w_{x_2} + \frac{\bar u^{(3)}}{1+\bar u^{(1)}} w_{x_3} = \frac{v}{1+\bar u^{(1)}} & \textrm{in} & {\cal D'}(\Omega), \\
w = w_{in} & \textrm{on} & \Gamma_{in},
\end{array} \right.
\end{equation}
and, treating $x_1$ as a 'time' variable, adapt Di Perna - Lions theory of transport equation (\cite{DPL})
that implies the uniqueness of solution to (\ref{def_S_1}) in the class $L_{\infty}(L_2)$.
The proof is thus complete.
$\square$

\subsection{Strong solution}
Having the weak solution of the linear system (\ref{system_lin}) we can show quite easily that this solution is
strong if the data has the appropriate regularity. The following lemma gives existence of a strong
solution to (\ref{system_lin}).
\begin{lem}
Let $F \in L_p$, $G \in W^1_p$, $w_{in} \in W^1_p(\Gamma_{in})$, $B \in W^{1-1/p}_p(\Gamma)$
and assume that $f$ is large enough and $||\bar u||_{W^2_p}$ is small enough. Then there exist
$(u,w) \in W^2_p \times W^1_p$ that is a strong solution to (\ref{system_lin}) 
and satisfies the estimate (\ref{est_lin_w2p}).
\end{lem}
\textbf{Proof.} Since (\ref{system_lin}) is a linear system, the \emph{a priori} estimate
(\ref{est_lin_w2p}) will imply the regularity of the weak solution once we can deal with
the singularity of the boundary at the juctions of $\Gamma_0$ with $\Gamma_{in}$ and $\Gamma_{out}$.
This however can be done easily since $\Omega$ is symmetric w.r.t. the plane $\{ x_1=0 \}$
and the slip boundary conditions preserve this symmetry. More precisely, for 
$\{ \tilde x = (-x_1,x_2,x_3): x = (x_1,x_2,x_3) \in \Omega \}$ we can can consider a vector field 
\begin{equation} \label{reflected}
\tilde u(\tilde x) = [- u^1(x), u^2(x), u^3(x)].    
\end{equation}
Then on $\Gamma_{in}$ we have $\tilde u \cdot n = u \cdot n$ and 
$n \cdot {\bf D}(\tilde u) \cdot \tau_i + \tilde u \cdot \tau_i$ = $n \cdot {\bf D}(u) \cdot \tau_i + u \cdot \tau_i$. 
Hence we can extend the weak solution on the negative values of $x_1$ using (\ref{reflected}) and, applying
the estimate (\ref{est_lin_w2p}), show that the extended solution is in $W^2_p \times W^1_p$.
An identical argument can be applied on $\Gamma_{out}$ and we coclude that $(u,w)$
is a strong solution to (\ref{system_lin}). $\square$ 
\section{Bounds on the approximating sequence}  \label{sec_conv}
In this section we will show the bounds on the sequence $\{(u^n,w^n)\}$
of solutions to (\ref{system_seq}). The term $u \cdot \nabla w$ in the continuity equation makes it impossible
to show directly the convergence in $W^2_p \times W^1_p$ to the strong solution of (\ref{system}).
We can show however that the sequence of iterated solutions is bounded in $W^2_p \times W^1_p$, and using this
bound we can conclude it is a Cauchy sequence in $H^1 \times L_{\infty}(L_2)$, and thus converges in this space
to some couple $(u,w)$. On the other hand, the boundedness implies weak
convergence in $W^2_p \times W^1_p$, and the limit must be $(u,w)$. 
The following lemma gives the boundedness of $(u^n,w^n)$ in $W^2_p \times W^1_p$. 
\begin{lem}  \label{lem_seq_bound}
Let $\{(u^n,w^n)\}$ be a sequence of solutions to (\ref{system_seq}) starting from $(u^0,w^0)=(0,0)$. Then
\begin{equation}
||u^n||_{W^2_p} + ||w^n||_{W^1_p} \leq M,  \label{est_seq_bound}
\end{equation}
where $M$ can be arbitrarily small provided that $||u_0||_{W^2_p}$
(extension of the boundary data (\ref{extension}), not to be confused with $u^0$ from $(u^0,w^0)$,
the starting point of the sequence $(u^n,w^n)$), $||B||_{W^{1-1/p}_p(\Gamma)}$, $||w_{in}||_{W^1_p(\Gamma_{in})}$
and $||\bar u||_{W^2_p}$ are small enough and $f$ is large enough.
\end{lem}
\textbf{Proof.} The estimate (\ref{est_lin_w2p}) for the iterated system reads
\begin{equation}  \label{unwn_1}
\begin{array}{c}
||u^{n+1}||_{W^2_p} + ||w^{n+1}||_{W^1_p} \leq \\
\leq C \, \big[ ||F(u^n,w^n)||_{L_p} + ||G(u^n,w^n)||_{W^1_p} + ||B||_{W^{1-1/p}_p(\Gamma)} + ||w_{in}||_{W^1_p(\Gamma_{in})} \big].
\end{array}
\end{equation}
Denoting $A_n = ||u^n||_{W^2_p} + ||w^n||_{W^1_p}$ and $b = ||u_0||_{W^2_p} + ||B||_{W^{1-1/p}_p(\Gamma)} + ||w_{in}||_{W^1_p(\Gamma_{in})}$,
from (\ref{FGlp}) and (\ref{unwn_1}) we get
\begin{equation} \label{an1}
A_{n+1} \leq A_n^2 + b,
\end{equation}
thus $A_n$ is bounded by a constant that can be arbitrarily small
provided that $A_0$ and $b$ are small enough.
Indeed let us fix $0 < \delta < \frac{1}{4}$ and assume that $b<\delta$.
Then (\ref{an1}) entails an implication $A_n \leq 2b \Rightarrow A_{n+1} \leq 2b$
and we can conclude that
\begin{equation}
\left. \begin{array}{c}
\delta < \frac{1}{4} \\
b < \delta \\
A_0 < 2b
\end{array} \right\}
\Rightarrow A_n < 2 \delta \quad \forall \, n \in \mathbb{N}.
\end{equation}
Hence if we fix $0 < \epsilon < \frac{1}{4}$ and assume that 
$||u_0||_{W^2_p} + ||B||_{W^{1-1/p}_p(\Gamma)} + ||w_{in}||_{W^1_p(\Gamma_{in})} < \epsilon$
then starting the iteration from $(u^0,w^0) = (0,0)$ we have
\begin{equation}
||u^n||_{W^2_p} + ||w^n||_{W^1_p} \leq 2 \delta \quad \forall \, n \in \mathbb{N}. \; \square
\end{equation}

The next lemma almost completes the proof of the Cauchy condition in $H^1 \times L_{\infty}(L_2)$
for the sequence of iterated solutions.
\begin{lem} \label{lem_cauchy1}
Let the assumptions of Lemma \ref{lem_seq_bound} hold. Then we have
\begin{equation}  \label{lem_cauchy1_teza}
||u^{n+1}-u^{m+1}||_{H^1} + ||w^{n+1}-w^{m+1}||_{L_{\infty}(L_2)} \leq
E(M) \, \big( ||u^{n}-u^{m}||_{H^1} + ||w^{n}-w^{m}||_{L_{\infty}(L_2)} \big),
\end{equation}
where $M$ is the constant from (\ref{est_seq_bound}).
\end{lem}

\textbf{Proof.}
Subtracting (\ref{system_seq})$_m$ from (\ref{system_seq})$_n$ we arrive at
\begin{displaymath}
\begin{array}{c}
\partial_{x_1} (u^{n+1} - u^{m+1}) - \mu \Delta (u^{n+1} - u^{m+1}) - (\nu+\mu) \nabla \, {\rm div}\, (u^{n+1} - u^{m+1}) \\
+ \gamma \nabla (w^{n+1} - w^{m+1}) = F(u^n,w^n) - F(u^m,w^m),
\end{array}
\end{displaymath}
\begin{displaymath}
\begin{array}{c}
{\rm div}\, (u^{n+1}-u^{m+1}) + \partial_{x_1} (w^{n+1} - w^{m+1}) + (u^n+u_0) \cdot \nabla (w^{n+1}-w^{m+1}) = \\
= G(u^n,w^n) - G(u^m,w^m) + (u^n-u^m) \cdot \nabla w^m,
\end{array}
\end{displaymath}
\begin{displaymath}
\begin{array}{c}
n\cdot 2\mu {\bf D}(u^{n+1}-u^{m+1})\cdot \tau_i +f \, (u^{n+1}-u^{m+1}) \cdot \tau_i|_{\Gamma} = 0, \\
n\cdot  (u^{n+1}-u^{m+1})|_{\Gamma} = 0, \\
w^{n+1}-w^{m+1}|_{\Gamma_{in}}=0.
\end{array}
\end{displaymath}
The estimate (\ref{ene1}) applied to this system yields
\begin{displaymath}
\begin{array}{c}
||u^{n+1}-u^{m+1}||_{H^1} + ||w^{n+1}-w^{m+1}||_{L_{\infty}(L^2)} \leq \nonumber\\
||F(u^n,w^n) - F(u^m,w^m)||_{V^*} + ||G(u^n,w^n) - G(u^m,w^m)||_{L_2} + ||(u^n - u^m) \cdot \nabla w^m||_{L_2}.
\end{array}
\end{displaymath}
In order to derive (\ref{lem_cauchy1_teza}) from the above inequality we have to examine the l.h.s.
The part with $G$ is the most straighforward and we have
\begin{equation}
||G(u^n,w^n) - G(u^m,w^m)||_{L_2} \leq E(M) \, \big( ||u^n-u^m||_{H^1} + ||w^n-w^m||_{L_{\infty}(L_2)} \big).
\end{equation}
The function $F$ is more complicated and we have to look at the difference more carefully.
A direct calculation yields
$F(u^n,w^n) - F(u^m,w^m) = F^{n,m}_1 + F^{n,m}_2$, where
\begin{equation}
||F^{n,m}_1||_{V^*} \leq E(M) \, \big( ||u^n-u^m||_{H^1} + ||w^n-w^m||_{L_{\infty}(L_2)} \big)
\end{equation}
and
\begin{equation} \label{fnm2}
F^{n,m}_2 = - [\delta \pi'(w^n) - \delta \pi'(w^m)] \nabla w^n + \delta \pi'(w^m) \nabla (w^n-w^m)
=: F^{n,m}_{2,1} + F^{n,m}_{2,2},
\end{equation}
where $\delta \pi'(\cdot)$ is defined in (\ref{delta_pi}).
Since we are interested in the $V^*$-norm of $F^{n,m}_2$, we have to multiply
$F^{n,m}_{2,1}$ and $F^{n,m}_{2,2}$ by $v \in V$ and integrate. With $F^{n,m}_{2,2}$ we get
\begin{displaymath}
\begin{array}{c}
\int_{\Omega} \delta \pi'(w^m) \nabla (w^n-w^m) \cdot v \,dx  = \\
-\int_{\Omega} \delta \pi'(w^m) (w^n - w^m) \, {\rm div}\,v \,dx
- \int_{\Omega} (w^n - w^m) \, \nabla \, [\delta \pi'(w^m)] \cdot v \,dx,
\end{array}
\end{displaymath}
and thus we have to estimate $\delta \pi'(w^m)$ in terms of $w^m$.
Using (\ref{mean}) we can write
\begin{equation}  \label{delta_p_wm}
\delta \pi'(w^m) = w^m \, \int_{0}^1 \pi''[tw^m+1] \,dt,
\end{equation}
what yields
$
||\delta \pi'(w^m)||_{L_{\infty}} \leq C(\pi) ||w^m||_{L_{\infty}}.
$
Now we have to estimate $||\nabla \delta \pi'(w^m)||_{L_p}$. Since $\pi$ is a $C^3$ - function
(and this is the only point where $C^3$ - regularity is needed) we can take
the gradient of (\ref{delta_p_wm}) and verify that
$
||\nabla \delta \pi'(w^m)||_{L_p} \leq C(\pi) ||\nabla w^m||_{L_p}.
$
Thus we have
\begin{equation}  \label{fnm22_1}
\begin{array}{c}
\big| \int_{\Omega} \delta \pi'(w^m) (w^n - w^m) \, {\rm div}\,v \,dx \big|
\leq ||\delta \pi'(w^m)||_{L_{\infty}} \, ||w^n-w^m||_{L_2} \, ||{\rm div}\,v||_{L_2} \leq \\
\leq C\, ||w^m||_{W^1_p} \, ||w^n-w^m||_{L_{\infty}(L_2)} \, ||v||_{V}.
\end{array}
\end{equation}
Next, since $p>3$, by the Sobolev imbedding theorem we have
\begin{equation}  \label{fnm22_2}
\begin{array}{c}
\big| \int_{\Omega} (w^n - w^m) \, \nabla \, [\delta \pi'(w^m)] \cdot v \,dx \big| \leq  \\
\leq ||w^n - w^m||_{L_2} \, ||\nabla \delta \pi'(w^m)||_{L_p} \, ||v||_{L_6} \leq
C\, ||w^m||_{W^1_p} \, ||w^n - w^m||_{L_{\infty}(L_2)} \, ||v||_{V}
\end{array}
\end{equation}
Combining (\ref{fnm22_1}) and (\ref{fnm22_2}) we get
\begin{equation}
||F^{n,m}_{2,2}||_{V^*} \leq E(M) \, ||w^n-w^m||_{L_{\infty}(L_2)}.
\end{equation}
In order to estimate $F^{n,m}_{2,1}$ we will use again $(\ref{mean})$ to write
\begin{equation}
\delta \pi'(w^n) - \delta \pi'(w^m) = (w^n-w^m) \, \int_0^1 p''[t\,w^n + (1-t) \,w^m +1 ] \,dt,
\end{equation}
what yields $||\delta \pi'(w^n) - \delta \pi'(w^m)||_{L_2} \leq C \, ||w^n-w^m||_{L_2}$.
With this observation we can estimate
\begin{displaymath}
\begin{array}{c}
\big| \int_{\Omega} [ \delta \pi'(w^n) - \delta \pi'(w^m) ] \nabla w^n \cdot v \,dx \big| \leq
||\delta \pi'(w^n) - \delta \pi'(w^m)||_{L_2} \, ||\nabla w^n||_{L_p} \, ||v||_{L_6} \leq \\
\leq E(||w^n||_{W^1_p}) \, ||w^n - w^m||_{L_{\infty}(L_2)} \, ||v||_{V},
\end{array}
\end{displaymath}
what yields
\begin{equation}
||F^{n,m}_{2,1}||_{V^*} \leq E(M) \, ||w^n-w^m||_{L_{\infty}(L_2)}.
\end{equation}
Combining the estimates on $F^{n,m}_1$,$F^{n,m}_{2,1}$ and $F^{n,m}_{2,2}$ we get
\begin{equation} \label{est_fnm}
||F(u^n,w^n)-F(u^m,w^m)||_{V^*} \leq E(M) \, [||u^n-w^n||_{H^1} + ||w^n-w^m||_{L_{\infty}(L_2)}].
\end{equation}
The part that remains to estimate is $(u^n-u^m) \cdot \nabla w^m$.
We shall notice here that this is the term which makes it impossible to show the convergence
in $W^2_p \times W^1_p$ directly. Namely, if we would like to apply the estimate (\ref{est_lin_w2p})
to the system for the difference then we would have to estimate $||(u^n-u^m) \cdot \nabla w^m||_{W^1_p}$
what can not be done as we do not have any knowledge about $||w||_{W^2_p}$.

Fortunately we only need to estimate the $L_2$-norm of this awkward term, what is straightforward.
Namely, we have
\begin{equation}
||(u^n-u^m) \cdot \nabla w^m||_{L_2} \leq ||u^n-u^m||_{L_q} \, ||\nabla w^m||_{L_p}
\leq C \, ||w^m||_{W^1_p} \, ||u^n-u^m||_{H^1},
\end{equation}
since $q = \frac{2p}{p-2} <6$ for $p<3$. We have thus completed the proof of (\ref{lem_cauchy1_teza}). $\square$

Now, lemma \ref{lem_seq_bound} implies that the constant $E(M) < 1$ provided that
the data is small enough and the starting point $(u^0,w^0) = (0,0)$.
It completes the proof of the Cauchy condition in $H^1 \times L_{\infty}(L_2)$ for the sequence $\{(u^n,w^n)\}$.

{\bf Remark.} Lemmas \ref{lem_seq_bound} and \ref{lem_cauchy1} hold for any starting point $(u^0,w^0)$
small enough in $W^2_p \times W^1_p$, not necessarily $(0,0)$, but we can start the iteration from $(0,0)$
without loss of generality.
\section{Proof of Theorem \ref{main}} \label{sec_proof}
In this section we prove our main result, Theorem \ref{main}. First we show existence of the solution
passing to the limit with the sequence $(u^n,w^n)$ and next we show that this solution is unique
in the class of solutions satisfying (\ref{est_main}).

{\bf Existence of the solution}.
Since we have the Cauchy condition on the sequence $(u^n,w^n)$ only in the space
$H^1(\Omega) \times L_{\infty}(L_2)$, first we have to show the convergence in the weak formulation
of the problem (\ref{system}), transfering the derivatives of the density on the
test function. The sequence $(u^n,w^n)$ satisfies in particular the
following weak formulation of (\ref{system_seq})
\begin{equation} \label{weak1_seq}
\begin{array}{c}
\int_{\Omega} \{ v \cdot \partial_{x_1} u^{n+1} + 2 \mu {\bf D}(u^{n+1}) : \nabla \,v + \nu \, {\rm div}\,u^{n+1} \, {\rm div}\,v
- \gamma \, w^{n+1} {\rm div}\,v \} \,dx \\
+ \int_{\Gamma} f (u^{n+1} \cdot \tau_i) \, (v \cdot \tau_i) \,d\sigma
= \int_{\Omega} F(u^n,w^n) \cdot v \,dx + \int_{\Gamma} B_i (v \cdot \tau_i) \,d\sigma
\end{array}
\end{equation}
and
\begin{equation} \label{weak2_seq}
\begin{array}{c}
-\int_{\Omega} w^{n+1} [\tilde u^n \cdot \nabla \phi + {\rm div}\, \tilde u^n \, \phi] \,dx + \int_{\Gamma_{out}} w^{n+1} \, \phi \,d\sigma = \\
= \int_{\Omega} \phi (G(u^n,w^n) - {\rm div}\, u^{n+1})\,dx + \int_{\Gamma_{in}} w_{in} \, \phi \,d\sigma
\end{array}
\end{equation}
$\forall (v,\phi) \in V \times \bar C^{\infty}(\Omega)$,
where
%$$
%\tilde F(u,w) = F(u,w) + [\pi'(w+1) - \pi'(1)] =
%$$$$
%= - w \, (u + \bar v + u_0) \cdot \nabla (u+u_0) - (u_0 \cdot \nabla u) - u \cdot \nabla u_0
%+ \mu \Delta u_0 + (\nu+\mu) \nabla {\rm div}\, u_0 - u_0 \cdot \nabla u_0
%$$
$\tilde u^n = [1+(u^n+u_0)^{(1)},(u^n+u_0)^{(2)},(u^n+u_0)^{(3)}]$.

Now using the convergence in $H^1 \times L_{\infty}(L_2)$ combined with the bound (\ref{est_seq_bound})
in $W^2_p \times W^1_p$ we can pass to the limit in (\ref{weak1_seq}) - (\ref{weak2_seq}).
The convergence in all the terms on the r.h.s. of (\ref{weak1_seq}) is obvious and the only nontrivial step to show the
convergence of $F(u^n,w^n)$ is to show that 
$$
\int_{\Omega} \delta \pi'(w^n) \nabla w^n \cdot v \,dx \to  \int_{\Omega} \delta \pi'(w) \nabla w \cdot v \,dx.
$$
To show the above convergence it is enough to verify that
\begin{equation} \label{conv_1}
\int_{\Omega} [w^n \delta \pi'(w^n) - w \delta \pi'(w)] \, {\rm div} \, v \,dx \to 0
\end{equation}
and
\begin{equation} \label{conv_2}
\int_{\Omega} [w^n \nabla \delta \pi'(w^n) - w \nabla \delta \pi'(w)] \cdot v \,dx \to 0.
\end{equation}
Applying again (\ref{mean}) we have 
$\delta \pi'(w^n) - \delta \pi'(w) = (w^n - w) \int_0^1 \pi''( 1+ tw^n + (1-t)w ) \,dt$,
hence
\begin{equation} \label{est_dif_delta}
||\delta \pi'(w^n) - \delta \pi'(w)||_{L_2} \leq C \, ||w^n - w||_{L_2}, 
\end{equation}
what implies directly (\ref{conv_1}). To show (\ref{conv_2}) we integrate by parts arriving at
$$
\int_{\Omega} [w^n \nabla \delta \pi'(w^n) - w \nabla \delta \pi'(w)] \cdot v \,dx
= - \int_{\Omega} w^n \, (\delta \pi'(w^n) - \delta \pi'(w)) \, {\rm div} \, v \,dx
$$$$
 - \int_{\Omega} (\delta \pi'(w^n) - \delta \pi'(w)) \, v \cdot \nabla w^n \,dx
+ \int_{\Omega} (w^n - w) \, \nabla \delta \pi'(w) \cdot v \,dx.
$$
Now all the terms on the l.h.s. converge by (\ref{est_dif_delta}), hence we have shown (\ref{conv_2}).
 
We conclude that $(u,w)$ satisfies
\begin{equation} \label{weak1_lim}
\begin{array}{c}
\int_{\Omega} \{ v \cdot \partial_{x_1} u + 2 \mu {\bf D}(u) : \nabla \,v + \nu \, {\rm div}\,u \, {\rm div}\,v
- \gamma \, w \, {\rm div}\,v \} \,dx \\
+ \int_{\Gamma} f (u \cdot \tau_i) \, (v \cdot \tau_i) \,d\sigma
= \int_{\Omega} F(u,w) \cdot v \,dx + \int_{\Gamma} B_i(v \cdot \tau_i) \,d\sigma
\end{array}
\end{equation}
$\forall \; v \in V$. In $(\ref{weak2_seq})$ we have to check the convergence in the boundary term. We can use the same argument
as in the proof of the existence of solution to the linear system when we have passed to the limit with
finite dimensional approximations. Namely, in fact $w^n$ satisfies the Cauchy condition not only in
$L_\infty(L_2)$. A stronger fact holds that $w^n$ is a Cauchy sequence in $L_2(\Omega_{x_1})$
for every $x_1 \in [0,L]$, where $\Omega_{x_1}$ denotes the $x_1$-cut of $\Omega$.
In particular $w^n \to \zeta$ in $L_2(\Gamma_{out})$ for some $\zeta \in L_2(\Gamma_{out})$ and
since ${\rm sup}_{x_1 \in [0,L]} ||w||_{L_2(\Omega_{x_1})} < \infty$ we conclude that
$\zeta = w|_{\Gamma_{out}}$. This result combined with the obvious convergence of other terms in
(\ref{weak2_seq}) implies

\begin{equation} \label{weak2_lim}
-\int_{\Omega} w [\tilde u \cdot \nabla \phi + {\rm div}\, \tilde u \phi ] \,dx + \int_{\Gamma_{out}} w \, \phi \,d\sigma  =
\int_{\Omega} \phi (G(u,w) - {\rm div}\, u) \,dx + \int_{\Gamma_{in}} w_{in} \, \phi \,d\sigma
\end{equation}
$\forall \phi \in \bar C^{\infty}(\Omega)$, where $\tilde u = [1+(u+u_0)^{(1)},(u+u_0)^{(2)},(u+u_0)^{(3)}]$.

Hence we have shown that $(u,w)$ satisfies (\ref{weak1_lim}) - (\ref{weak2_lim}), the weak formulation of (\ref{system}).
Now we want to show that the strong formulation also holds.

The bound in $W^2_p \times W^1_p$ implies $(u^{n_k},w^{n_k}) \rightharpoonup (\bar u, \bar w)$
in $W^2_p \times W^1_p$ for some $(\bar u , \bar w) \in W^2_p \times W^1_p$.
On the other hand, we have $(u^{n_k},w^{n_k}) \to (u,w)$ in $H^1 \times L_{\infty}(L_2)$, thus we conclude
that $(\bar u, \bar w) = (u,w)$.

Hence we can integrate by parts in (\ref{weak1_lim}) - (\ref{weak2_lim}) to obtain
\begin{equation}
\begin{array}{c}
\int_{\Omega} \big[ F(u,w) - \mu \Delta u - (\mu+\nu) \nabla {\rm div}\,u + \gamma \nabla w \big] \cdot v \,dx \\
= \int_{\Gamma} \big[ B_i(v \cdot \tau_i) - n \cdot [ 2\mu \mathbf{D}(u) + \nu {\rm div} \,u \, {\bf Id}] \cdot v  -  f(u \cdot \tau_i)(v \cdot \tau_i) \big] \, d\sigma
\end{array}
\end{equation}
and
\begin{equation}
\int_{\Omega} [w_{x_1} + (u+u_0) \cdot \nabla w] \phi \,dx = \int_{\Omega} [G(u,w) - {\rm div}\,u] \phi \,dx.
\end{equation}
From these equations we conclude that (\ref{system})$_{1,2}$ are satisfied a.e. in $\Omega$
and (\ref{system})$_3$ is satisfied a.e. on $\Gamma$.
It remains to verify that (\ref{system})$_4$
is satisfied a.e. on $\Gamma$ and (\ref{system})$_5$ holds a.e. on $\Gamma_{in}$.
The condition (\ref{system})$_4$ results from the convergence $u^n \to u$ in $H^1$.

Finally, $w^n \rightharpoonup w$ in $W^1_p$ implies that
$w^n|_{\Gamma_{in}} \rightharpoonup tr\,w|_{\Gamma_{in}}$ in $L_p(\Gamma_{in})$.
On the other hand $w^n|_{\Gamma_{in}} \to w_{in}$ in $W^1_p(\Gamma_{in})$ since it is a constant sequence.
We conclude that $w|_{\Gamma_{in}} = w_{in}$.

{\bf Uniqueness.} In order to prove the uniqueness of the solution
%we can follow step by step the proof
%of the uniqueness in two dimensional case from \cite{TP}, thus we only sketch the proof here
%and refer to the details to \cite{TP}.
consider $(v_1,\rho_1)$ and $(v_2,\rho_2)$ being two solutions
to (\ref{main_system}) satisfying (\ref{est_main}). We will prove that
\begin{equation} \label{est_dif0}
||v_1 - v_2||_{H^1}^2 + ||\rho_1 - \rho_2||_{L_2}^2 = 0.
\end{equation}
For simplicity let us denote $u := v_1 - v_2$ and $w := \rho_1 - \rho_2$.
We will show that
\begin{equation} \label{est_dif1}
||u||_{H^1} \leq E ||w||_{L_2}
\end{equation}
and
\begin{equation} \label{est_dif2}
||w||_{L_2} \leq C ||u||_{H^1},
\end{equation}
what obviously implies (\ref{est_dif0}).
Subtracting the equations (\ref{main_system}) for $(v_1,\rho_1)$ and $(v_2,\rho_2)$
we get
\begin{equation}  \label{dif}
\begin{array}{l}
w \, v_2 \cdot \nabla v_2 + \rho_1 \, u \cdot \nabla v_2 + \rho_1 \, v_1 \cdot \nabla u
- \mu \Delta u - (\mu+\nu) \nabla {\rm div}\, u + I_\pi \nabla w + w \nabla I_\pi = 0,  \\
\rho_1 \, {\rm div}\, u + w \, {\rm div}\, v_2 + u \cdot \nabla \rho_2 + v_1 \cdot \nabla w = 0, \\
n \cdot 2 \mu \mathbf{D}(u) \cdot \tau|_{\Gamma} = 0, \\
n \cdot u|_{\Gamma} = 0, \\
w|_{\Gamma_{in}} = 0,
\end{array}
\end{equation}
where
\begin{equation} \label{i_pi}
I_\pi = \int_0^1 \pi'((t \rho_1) + (1-t) \rho_2)\,dt.
\end{equation}
Notice that $I_\pi \in W^1_p$ since $\rho_i \in W^1_p$ and $\pi \in C^3$.
In order to show (\ref{est_dif1}) we follow the proof of (\ref{ene1})
multiplying (\ref{dif})$_1$ by $\rho_1 \, u$ (it will be clarified soon why take the test function
$\rho_1 \, u$ instead of $u$). Using (\ref{basic_id}) we get
$$
\int_\Omega (2 \mu \mathbf{D^2}(u) + \nu \rho_1 \, div^2 \, u)\,dx
+ \underbrace{ \int_\Omega \Big\{ 2\mu \big[ (\rho_1-1) \mathbf{D}(u) : \nabla u + \mathbf{D}(u) : (u \otimes \nabla \rho_1) \big] + \nu ({\rm div} \,u) u \cdot \nabla \rho_1 \Big\} \,dx  }_{I_1}
$$$$
- \underbrace{ \int_{\Omega} \Big\{ w \, u \, \nabla \rho_1 +
 \rho_1^2 u^2 \cdot \nabla v_2 +
 u \, w \, \rho_1 \, v_2 \cdot \nabla v_2 \Big\} \,dx  }_{I_2}
+ \underbrace{ \int_{\Omega}  \rho_1^2 \, (v_1 \cdot \nabla u) \cdot u \,dx }_{I_3}
$$$$
+ \underbrace{ \int_{\Omega} \rho_1 \, w \, u \cdot \nabla I_{\pi} \,dx }_{I_4}
- \underbrace{ \int_{\Omega} w u \cdot \nabla (I_\pi \rho_1) \,dx }_{I_5}
- \int_\Omega I_{\pi} w \, \rho_1 \, {\rm div} \, u \,dx
+ \int_{\Gamma} \rho_1 \, f \, u^2 \, d\sigma = 0.
$$
We have $|I_1| + |I_2| \leq E \, (||u||_{H^1}^2 + ||w||_{L_2}^2 )$ and in order to deal with $I_3$
let us split it into two parts:
$$
2 I_3 = \underbrace{ \int_{\Omega} \big\{ (\rho_1^2 \, v_1^{(1)} - 1) \,
\partial_{x_1} |u|^2 + \rho_1^2 \, v_1^{(2)} \, \partial_{x_2} |u|^2 + \rho_1^2 \, v_1^{(3)} \, \partial_{x_3} |u|^2 \big\} \,dx}_{I_3^1}
+ \underbrace{ \int_{\Omega} \partial_{x_1} |u|^2 \,dx}_{I_3^2}.
$$
We have
$|I_3^1| \leq E ||u||_{H^1}^2$
and
$
I_3^2 = \int_{\Gamma} |u|^2 n^{(1)} \, d\sigma =
- \int_{\Gamma_{in}} |u|^2 \, d\sigma + \int_{\Gamma_{out}} |u|^2 \, d\sigma.
$
In order to examine $I_4$ and $I_5$ we have to differentiate (\ref{i_pi}) what yields
\begin{equation} \label{nabla_ipi}
\nabla I_\pi = I_\pi^1 \nabla \rho_1  + I_\pi^2 \nabla \rho_2 ,
\end{equation}
where
$$
I_\pi^1 = \int_0^1 \pi''(t\rho_1 + (1-t)\rho_2)t \,dt \quad {\rm and} \quad
I_\pi^2 = \int_0^1 \pi''(t\rho_1 + (1-t)\rho_2)(1-t) \,dt.
$$
We have
$$
| \int_{\Omega} \rho_1 \, I_\pi^1 \, u \, w \, \nabla \rho_1 \,dx | \leq
||\rho_1 \, I_\pi^1||_{L_\infty} \, ||\nabla \rho_1||_{L_p} \, ||u||_{L_6} \, ||w||_{L_2}
\leq E \, (||u||_{H^1}^2 + ||w||_{L_2}^2),
$$
and the same for $\int_{\Omega} \rho_1 \, I_\pi^2 \, u \, w \, \nabla \rho_2 \,dx$.
Thus the application of (\ref{nabla_ipi}) to $I_4$
yields $|I_4| \leq E \, (||u||_{H^1}^2 + ||w||_{L_2}^2)$.
To estimate $|I_5|$ it is enough to use (\ref{nabla_ipi})
to compute $\nabla (I_{\pi} \rho_1)$ and then with the same arguments as in case of $I_4$
we get $|I_5| \leq E \, (||u||_{H^1}^2 + ||w||_{L_2}^2 )$.
Summarizing our estimates we can write
\begin{equation} \label{est_dif1_1}
\begin{array}{c}
||u||_{H^1}^2 + \int_{\Gamma_{in}} (\rho_1 \, f - \frac{1}{2})|u|^2 \,d\sigma
+ \int_{\Gamma_0} \rho_1 \, f \,|u|^2 \,d\sigma + \int_{\Gamma_{out}} (\rho_1 \, f + \frac{1}{2})|u|^2 \,d\sigma \leq
\\
\leq \int_{\Omega} I_\pi \, w \, \rho_1 {\rm div} \,u \,dx + E \, ||w||_{L_2}^2.
\end{array}
\end{equation}
The boundary integrals over $\Gamma_0$ and $\Gamma_{out}$ will be nonnegative for any $f \geq 0$
and the integral over $\Gamma_{in}$ will be nonnegative for $f$ large enough on $\Gamma_{in}$.
Now in order to obtain (\ref{est_dif1}) from (\ref{est_dif1_1}) we can express $\rho_1 \, {\rm div \,u}$
in terms of $w$ using the equation (\ref{dif})$_2$ 
(this is why we have tested (\ref{dif})$_1$ with $\rho_1 \, u$ instead of $u$) and rewrite (\ref{est_dif1_1}) as
\begin{equation}
||u||_{H^1}^2 \leq - \underbrace{ \int_{\Omega} I_\pi \, w^2 \, {\rm div}\, v_2 \,dx }_{I_6}
- \underbrace{ \int_{\Omega} I_\pi \, w \, u \cdot \nabla \rho_2 \,dx }_{I_7}
- \underbrace{ \int_{\Omega} I_\pi \, v_1 \, w \cdot \nabla w \, dx }_{I_8}
+ E \, ||w||_{L_2}^2.
\end{equation}
We verify easily that $|I_6| + |I_7| \leq E \, (||u||_{H_1}^2 + ||w||_{L_2}^2)$.
We have to put a little more effort to find a bound on $I_8$. Let us integrate by parts:
$$
2 I_8 = \int_{\Omega} I_\pi \, v_1 \,\nabla w^2 \,dx =
- \int_{\Omega} w^2 {\rm div} (I_\pi \, v_1) \,dx + \int_{\Gamma} w^2 I_\pi v_1 \cdot n \,d\sigma.
$$
The boundary term reduces to $\int_{\Gamma_{out}} I_{\pi} w^2 v_1^{(1)} \,d\sigma > 0$ and in order to deal
with the first term on the l.h.s. notice that
$$
{\rm div} (I_\pi \, v_1) = {\rm div} v_1 \, I_\pi
+ I_\pi^1 \, v_1 \cdot \nabla \rho_1  + I_\pi^2 \, v_1 \cdot \nabla \rho_2,
$$
hence
$$
2 I_8 \leq - \underbrace{ \int_{\Omega} w^2 {\rm div} v_1 \, I_\pi \,dx }_{I_8^1}
- \underbrace{ \int_{\Omega} w^2 \, v_1 \cdot \nabla \rho_1 \, I_\pi^1 \,dx }_{I_8^2}
- \underbrace{ \int_{\Omega} w^2 \, v_1 \cdot \nabla \rho_2 \, I_\pi^2 \,dx }_{I_8^3}.
$$
Obviously we have $|I_8^1| \leq E \, ||w||_{L_2}^2$. In order to bound $I_8^2$ we can apply
the continuity equation that yields $v_i \cdot \nabla \rho_i = - \rho_i \, {\rm div} \,v_i$,
what implies
$
|I_8^2| = | \int_{\Omega} w^2 \, \rho_1 \, {\rm div} \,v_i \, I_\pi^1 \,dx | \leq E \, ||w||_{L_2}^2.
$
In the term $I_8^3$ we can rewrite the mixed component as
$v_1 \cdot \nabla \rho_2 = u \cdot \nabla \rho_2 + v_2 \cdot \nabla \rho_2$ and conclude that
$
|I_8^3| \leq E \, (||u||_{H^1}^2 + ||w||_{L_2}^2).
$
Combining the above results with (\ref{est_dif1_1}) we get (\ref{est_dif1}).

In order to show (\ref{est_dif1}) we express the pointwise value of $w$ using (\ref{dif})$_2$:
$$
w^2(x_1,x_2) = \int_0^{x_1} w \, w_s (s,x_2) ds =
- \int_0^{x_1} \frac{\rho_1}{v_1^{(1)}} w \, {\rm div}\,u (s,x_2) ds
$$$$
- \int_0^{x_1} \frac{1}{v_1^{(1)}}  \Big( w^2 \, {\rm div}\,v_2 + w \, u \cdot \nabla \rho_2 \Big) (s,x_2) ds
- \frac{1}{2} \int_0^{x_1} \frac{1}{v_1^{(1)}} \big[ v_1^{(2)} \partial_{x_2} w^2 + v_1^{(3)} \partial_{x_3} w^2 \big] (s,x_2) ds 
$$$$
=: w_1^2 + w_2^2 + w_3^2.
$$
We estimate directly the first two components of the l.h.s. obtaining
$$
\int_{\Omega} w_1^2 \,dx \leq \epsilon ||w||_{L_2}^2 + C(\epsilon) ||u||_{H_1}^2 \quad \forall \epsilon>0
$$
and
$
\int_{\Omega} w_2^2 \,dx \leq E \, (||w||_{L_2}^2 + ||u||_{H^1}^2).
$
To complete the proof we have to find a bound on $w_3^2$. To this end
notice that
$$
\int_{\Omega} w_3^2 \,dx = \frac{1}{2} \int_0^L \int_{P_{x_1}} \frac{1}{v_1^{(1)}} \big[ v_1^{(2)} \partial_{x_2} w^2 + v_1^{(3)} \partial_{x_3} w^2 \big] \,dx \,dx_1,
$$
where $P_{x_1} = \Omega_0 \times (0,x_1)$.
Integrating by parts in the inner integral we get
$$
\int_{\Omega} w_3^2 \,dx = \frac{1}{2} \int_0^L \Big\{ - \int_{P_{x_1}} w^2 \big[ \partial_{x_2} \frac{v_1^{(2)}}{v_1^{(1)}} + \partial_{x_3} \frac{v_1^{(3)}}{v_1^{(1)}} \big] \,dx
+ \int_{\partial P_{x_1}} \frac{w^2}{v_1^{(1)}} \big[ v_1^{(2)} n^{(2)} + v_1^{(3)} n^{(3)} \big] d\sigma \Big\} \,dx_1 .
$$
The boundary integral reduces to
$
\int_{\Gamma_0 \cap \partial P_{x_1}} w^2 \, v \cdot n \,d\sigma = 0,
$
what implies
$
\int_{\Omega} w^3_3 \,dx \leq E \, ||w||_{L_2}^2
$
and (\ref{est_dif2}) easily follows completing the proof of the uniqueness,
and hence the proof of the Theorem. $\square$

\section{Appendix}
\textbf{Vorticity on the boundary.} In order to show the boundary relation (\ref{system_rot})$_{3,4}$
we have to differentiate (\ref{system_lin})$_4$ with respect to tangential directions at a given point
$x_0 \in \Gamma$. Without loss of generality we can assume that $n(x_0)=(1,0,0)$, $\tau_1(x_0)=(0,1,0)$
and $\tau_2(x_0)=(0,0,1)$. Then we can rewrite (\ref{system_lin})$_3$ as (all the quantities are taken
at $x_0$):
\begin{equation}  \label{rot_bdry_1}
\left\{ \begin{array}{c}
\mu (u^1,_2 + u^2,_1) + f \, u^2 = B_1, \\
\mu (u^1,_3 + u^3,_1) + f \, u^3 = B_2.
\end{array} \right.
\end{equation}
Differantiating (\ref{system_lin})$_4$ with respect to the tangential direction $\tau_1$ we get
\begin{equation} \label{rot_bdry_2}
(\frac{d}{d \, \tau_1}n) \cdot u + u^1,_2 = 0.
\end{equation}
If we denote by $\chi_1$ the curvature of the curve generated by $\tau_1$ then we have
$\frac{d}{d \, \tau_1}n = \chi_1 \tau_1$ and (\ref{rot_bdry_2}) can be rewritten as
$
\chi_1 (\tau_1 \cdot u) + u^1,_2 = 0.
$
Combining this equation with (\ref{rot_bdry_1})$_1$ we get
$$
u^2,_1 - u^1,_2 = (2 \chi_1 - \frac{f}{\mu}) \, (u \cdot \tau_1) + \frac{B_1}{\mu},
$$
what is exactly (\ref{system_rot})$_3$. (\ref{system_rot})$_4$ can be shown in the same way
differentiating (\ref{system_lin})$_4$ with respect to the tangential direction $\tau_2$.
%
%\begin{lem} (Korn inequality): \label{lem_Korn}
%Assume that the friction coefficient $f$ is large enough. Then for $u \in V$:
%\begin{equation} \label{Korn}
%\int_{\Omega} 2 \mu {\bf D}^2(u) + \int_{\Gamma} f (u \cdot \tau)^2 \,d\sigma \geq C \, ||u||_{H^1}^2.
%\end{equation}
%\end{lem}
%{\bf Proof.} The condition $u \cdot n|_{\Gamma} = 0$ implies the Poincare inequality in $\Omega$ and hence
%we can repeat the proof of Lemma 2.4 in \cite{TP1}. $\square$

\begin{lem} (interpolation inequality): \label{lem_int} \\
$\forall \epsilon >0 \quad \exists C(\epsilon,p,\Omega)$ such that $\forall f \in W^1_p(Q)$:
\begin{equation}  \label{int1}
||f||_{L_p} \leq \epsilon ||\nabla f||_{L_p} + C \, ||f||_{L_2}.
\end{equation}
\end{lem}
\textbf{Proof.} Inequality (\ref{int1}) results from the inequality
$||f||_{L_p} \leq C(p,\Omega) \, ||f||_{W^1_2}^{\theta} \, ||f||_{L_2}^{1-\theta}$
for $2 \leq p < \infty$, where $\theta = \frac{n(p-2)}{2p}$
(see \cite{Ad}, Theorem 5.8).
Using Cauchy inequality with $\epsilon$ we get \ref{int1}.
$\square$

The last auxiliary result we use is a following fact on finitely dimensional Hilbert spaces
(the proof can be found in \cite{Te}):
\begin{lem}  \label{lem_P}
Let $X$ be a finite dimensional Hilbert space and let $P:X \to X$ be a continuous operator
satisfying
\begin{equation}  \label{lem_P_1}
\exists M>0: \quad (P(\xi),\xi) > 0 \quad \textrm{for} \quad ||\xi|| = M.
\end{equation}
Then
$
\exists \xi^*: \quad ||\xi^*|| \leq M \quad \textrm{and} \quad P(\xi^*) = 0.
$
\end{lem}

{\bf Acknowledgements.} The author would like to thank Piotr Mucha from the University of Warsaw
 for valuable comments and remarks concerning the paper.
 The author has been supported by the Polish Ministry of Science and Higher Education
grant No. N N201 364736.


\begin{thebibliography}{99}
\footnotesize
\bibitem{Ad} R.Adams, J.Fournier, \emph{Sobolev spaces}\/, 2nd ed., Elsevier, Amsterdam, 2003
\bibitem{ADN1} S.Agmon, A.Douglis, L.Nirenberg, \emph{Estimates near the boundary for solutions of elliptic
			partial differential equations satisfying general boundary conditions I}\/,
			Comm.Pure Appl.Math. 12 (1959), 623-727
\bibitem{ADN2} S.Agmon, A.Douglis, L.Nirenberg, \emph{Estimates near the boundary for solutions of elliptic
			partial differential equations satisfying general boundary conditions II}\/,
			Comm.Pure Appl.Math. 17 (1964), 35-92
\bibitem{MD} R.Danchin, P.B.Mucha, \emph{A critical functional framework for the inhomogeneous
			Navier-Stokes equations in the half-space}\/,  J. Funct. Anal.  256,3 (2009), 881-927
\bibitem{DPL} R.J.DiPerna, P.L.Lions, \emph{Ordinary differential equations, transport theory and Sobolev spaces}\/,
			Invent.math. 98 (1989), 511-547
\bibitem{Fe} E.Feireisl, \emph{Dynamics of viscous compressible fluids}\/,
			Oxford Lecture Series in Mathematics and its Applications, 26.
			Oxford University Press, Oxford, 2004
\bibitem{Ga} G.P.Galdi, \emph{An Introduction to the mathematical theory of the Navier-Stokes Equations}\/,
				Vol.I, Springer-Verlag, New York, 1994
\bibitem{Kw1} R.B.Kellogg, J.R.Kweon, \emph{Compressible Navier-Stokes equations in a bounded domain
				with inflow boundary condition}\/, SIAM J.Math.Anal. 28,1(1997), 94-108
\bibitem{Kw2} R.B.Kellogg, J.R.Kweon, \emph{Smooth Solution of the Compressible Navier-Stokes Equations
				in an Unbounded Domain with Inflow Boundary Condition}\/,
				J.Math.Anal. and App. 220 (1998), 657-675
\bibitem{Kw3} J.R.Kweon, M.Song, \emph{Boundary geometry and regularity of solution to the compressible
				Navier-Stokes equations in bounded domains of $\mathbf{R^n}$}\/,
				ZAMM Z.Angew.Math.Mech. 86,6 (2006), 495-504
\bibitem{PL} P.L.Lions, \emph{Mathematical topics in fluid mechanics. Vol. 2. Compressible models}\/,
			Oxford Lecture Series in Mathematics and its Applications, 10.
			Oxford Science Publications. The Clarendon Press, Oxford University Press, New York, 1998
\bibitem{PM1} P.B.Mucha, \emph{On cylindrical symmetric flows through pipe-like domains}\/,
			J.Differential Equations 201 (2004), 304-323
\bibitem{PM2} P.B.Mucha, \emph{On Navier-Stokes equations with Slip Boundary Conditions
			in an Infinite Pipe}\/, Acta Applicandae Mathematicae 76 (2003), 1-15
\bibitem{PMMP1} P.B.Mucha, M.Pokorny, \emph{On a new approach to the issue of existence and regularity
			for the steady compressible Navier-Stokes equations}\/, Nonlinearity 19(2006), 1747-1768
\bibitem{MR} P.B.Mucha, R.Rautmann, \emph{Convergence of Rothe's scheme for the Navier-Stokes equations with slip conditions in 2D domains}\/,
			ZAMM Z. Angew. Math. Mech. 86,9 (2006), 691-701
\bibitem{NoS} A.Novotny, I.Straskraba, \emph{An Introduction to the Mathematical Theory of Compressible Flows}\/,
			Oxford Science Publications, Oxford 2004
\bibitem{TP1} T.Piasecki, \emph{Steady Compressible Oseen Flow with slip boundary conditions}\/,
			Nonlocal and abstract parabolic equations and their applications,
			Banach Center Publications, Vol. 86, Institute of Matematics, Polish Academy of Sciences, Warsaw 2009
\bibitem{TP2} T.Piasecki, \emph{Steady compressible Navier-Stokes flow in a square},
			J.Math.Anal.Appl. 357 (2009), 447-467
\bibitem{PRS1} P.I.Plotnikov, E.V.Ruban, J.Sokolowski, \emph{Inhomogeneous boundary value problems for
			compressible Navier-Stokes Equations: well-posedness and sensitivity analysis}\/,
			SIAM J.Math.Anal. 40,3 (2008), 1152-1200
\bibitem{PRS2} P.I.Plotnikov, E.V.Ruban, J.Sokolowski,	\emph{Inhomogeneous boundary value problems
			for compressible Navier-Stokes an transport equations}\/,
			J.Math.Pures Appl. 92,2 (2009), 113-162
%\bibitem{PS1} P.I.Plotnikov, J.Sokolowski, \emph{Domain dependence of solutions to Compressible Navier-Stokes
%			Equations}\/, SIAM J.Control Optim. 45,4, 1165-1197
\bibitem{PS2} P.I.Plotnikov, J.Sokolowski, \emph{On Compactness, Domain Depedence and Existence of
			Steady State Solutions to Compressible Isothermal Navier-Stokes equations},
			J.Math.Fluid.Mech. 7 (2005), 529-573
\bibitem{PS3} P.I.Plotnikov, J.Sokolowski, \emph{Stationary Solutions of Navier-Stokes equations
				for diatomic gases}\/, Russian Math Surveys 62:3, 561-593
\bibitem{PMMP2} M.Pokorny, P.B.Mucha, \emph{3D Steady Compressible Navier-Stokes Equations}\/,
			Discrete and Continuous Dynamical Systems S, 1(1) (2008), 151-163
\bibitem{Te} R.Temam, \emph{Navier Stokes Equations}\/,	North-Holland, Amsterdam, 1977.
%\bibitem{Tr} H.Triebel, \emph{Interpolation theory, function spaces, differential operators.}\/,
%			North-Holland Mathematical Library, 18. North-Holland Publishing Co., Amsterdam-New York, 1978
\bibitem{VZ} A.Valli, W.M.Zajaczkowski, \emph{Navier - Stokes equations for compressible fluids: global existence and qualitative properties of the solutions in the general case}\/,
 			Comm. Math. Phys. 103,2 (1986), 259-296
\bibitem{Z} W.M.Zajaczkowski, \emph{Existence and regularity of solutions of some elliptic systems in domains
				with edges},\/ Dissertationes Math., 274(1989), 95 pp.
\end{thebibliography}
\end{document}